\documentclass{SciPost}

\hypersetup{
    colorlinks,
    linkcolor={red!50!black},
    citecolor={blue!50!black},
    urlcolor={blue!80!black}
}

\usepackage[bitstream-charter]{mathdesign}
\usepackage{enumitem}
\usepackage{algorithm}
\usepackage{algpseudocode}
\DeclareSymbolFont{usualmathcal}{OMS}{cmsy}{m}{n}
\DeclareSymbolFontAlphabet{\mathcal}{usualmathcal}

\fancypagestyle{SPstyle}{
\fancyhf{}
\lhead{\colorbox{scipostblue}{\bf \color{white} ~SciPost Physics }}
\rhead{{\bf \color{scipostdeepblue} ~Submission }}

\fancyfoot[C]{\textbf{\thepage}}
}

\newcommand{\be}{\begin{equation}}
\newcommand{\ee}{\end{equation}}
\newcommand{\bea}{\begin{eqnarray}}
\newcommand{\eea}{\end{eqnarray}}
\newcommand{\bmat}{\begin{pmatrix}}
\newcommand{\emat}{\end{pmatrix}}
\newcommand{\lb}{\left(}
\newcommand{\rb}{\right)}
\newcommand{\lsb}{\left[}
\newcommand{\rsb}{\right]}
\newcommand{\mc}{\mathcal}
\newcommand{\mf}{\mathfrak}
\newcommand{\mb}{\mathbf} 
\newcommand{\mr}{\mathrm}
\newcommand{\bs}{\boldsymbol}
\newcommand{\Tr}{\operatorname{Tr}}
\newcommand{\eff}{\mathrm{eff}}

\newcommand{\tb}[1]{\textcolor{blue}{#1}}

\newcommand{\old}[1]{\textcolor{gray}{#1}}

\begin{document}

\pagestyle{SPstyle}

\begin{center}{\Large \textbf{\color{scipostdeepblue}{
%%%%%%%%%% TODO: Write your article's title here
Magic of Kitaev spin liquids\\
%%%%%%%%%% END TODO: TITLE
}}}\end{center}

\begin{center}\textbf{
%%%%%%%%%% TODO: AUTHORS
% Write the author list here. 
% Use (full) first name (+ middle name initials) + surname format.
% Separate subsequent authors by a comma, omit comma and use "and" for the last author.
% Mark the corresponding author(s) with a superscript symbol in this order
% \star, \dagger, \ddagger, \circ, \S, \P, \parallel, ...
Eleonora Lamma\textsuperscript{1$*$},
Tim Bauer\textsuperscript{2},
Marcello Dalmonte\textsuperscript{2,3,4} and
Mario Collura\textsuperscript{1,5}
%%%%%%%%%% END TODO: AUTHORS
}\end{center}

\begin{center}
%%%%%%%%%% TODO: AFFILIATIONS
% Write all affiliations here.
% Format: institute, city, country
{\bf 1} International School for Advanced Studies (SISSA),\\ 
Via Bonomea 265, I-34136 Trieste, Italy
\\
{\bf 2} The Abdus Salam International Centre for Theoretical Physics (ICTP),\\
Strada Costiera 11, 34151 Trieste, Italy
\\
{\bf 3} INFN, Sezione di Bologna,\\ Via Irnerio 46, I-40126 Bologna, Italy\\
{\bf 4} Dipartimento di Fisica e Astronomia, Università di Bologna,\\ Via Irnerio 46, I-40126 Bologna, Italy\\ 
{\bf 5} INFN, Sezione di Trieste,\\
Via Valerio 2, 34127 Trieste, Italy\\
%%%%%%%%%% END TODO: AFFILIATIONS
%%%%%%%%%% TODO: EMAIL
% Provide email address of corresponding author(s)
%\\[\baselineskip]
$\star$ \href{mailto:elamma@sissa.it}{\small elamma@sissa.it}
%%%%%%%%%% END TODO: EMAIL
\end{center}

\section*{\color{scipostdeepblue}{Abstract}}
\textbf{\boldmath{%
%%%%%%%%%% TODO: ABSTRACT
% Write your abstract here.
Quantum spin liquids (QSLs) are long-range entangled phases of matter and a natural setting for exploring how quantum correlations %organize interacting many-body systems and 
generate quantum complexity.
%Beyond entanglement, magic, or nonstabilizerness, has recently emerged as a complementary diagnostic of the complexity of many-body states.}
%Quantum spin liquids (QSLs) are exotic phases of matter and a natural setting for exploring how genuinely quantum correlations organize strongly interacting many-body systems. 
%While long-range entanglement is a defining characteristic of QSLs, magic, or nonstabilizerness, has recently emerged as a complementary resource that characterizes the complexity of representing, preparing and simulating many-body quantum states. 
%Here, we investigate magic in QSLs by computing the stabilizer R{\'e}nyi entropy (SRE) of the Kitaev honeycomb model. 
Motivated by the emergence of magic, or nonstabilizerness, as a diagnostic of many-body complexity beyond entanglement, we study magic of QSLs by computing the stabilizer Rényi entropy (SRE) of the Kitaev honeycomb model.
We derive a correspondence between Pauli strings and products of itinerant Majorana operators in the model's free-fermion description; this enables the sampling of SRE using an optimized algorithm for Gaussian states in systems with thousands of spins. Our results show that magic is largest in the gapless phase, where subleading volume-law corrections indicate nonlocal contributions, while in the gapped phases, SRE decreases in quantitative agreement with a perturbative expansion that we develop in the anisotropic limit. %At the topological phase transition, we observe critical scaling of SRE dictated by critical exponents.
At the topological phase transition, SRE exhibits universal scaling dictated by critical exponents.
%Our results establish SRE as a probe of phase transitions between QSLs, and open new avenues for exploring the interplay between magic, long-range entanglement, and topological order.
%%%%%%%%%% END TODO: ABSTRACT
}}

\vspace{\baselineskip}

%%%%%%%%%% BLOCK: Copyright information
% This block will be filled during the proof stage, and finilized just before publication.
% It exists here only as a placeholder, and should not be modified by authors.
\noindent\textcolor{white!90!black}{%
\fbox{\parbox{0.975\linewidth}{%
\textcolor{white!40!black}{\begin{tabular}{lr}%
  \begin{minipage}{0.6\textwidth}%
    {\small Copyright attribution to authors. \newline
    This work is a submission to SciPost Physics. \newline
    License information to appear upon publication. \newline
    Publication information to appear upon publication.}
  \end{minipage} & \begin{minipage}{0.4\textwidth}
    {\small Received Date \newline Accepted Date \newline Published Date}%
  \end{minipage}
\end{tabular}}
}}
}
%%%%%%%%%% BLOCK: Copyright information

%%%%%%%%%% TODO: LINENO
% For convenience during refereeing we turn on line numbers:
%\linenumbers
% You should run LaTeX twice in order for the line numbers to appear.
%%%%%%%%%% END TODO: LINENO

%%%%%%%%%% TODO: TOC 
% Guideline: if your paper is longer that 6 pages, include a TOC
% To remove the TOC, simply cut the following block
\vspace{10pt}
\noindent\rule{\textwidth}{1pt}
\tableofcontents
\noindent\rule{\textwidth}{1pt}
\vspace{10pt}
%%%%%%%%%% END TODO: TOC

%%%%%%%%% TODO: CONTENTS 
% Write your article contents here, starting from first \section.
% An example structure is given below.

\section{Introduction}
    \label{sec:intro}

Quantum spin liquids (QSLs) continue to attract intense interest in both theoretical and experimental condensed matter physics, as they provide paradigmatic examples of phases that evade the conventional description in terms of local order parameters and spontaneous symmetry breaking \cite{savary2016quantum,broholm2020quantum}. 
Their low-energy physics is governed by intrinsically quantum features such as persistent quantum fluctuations, fractionalized excitations, emergent gauge fields, and, in certain cases, topological order~\cite{moessner2021topological}. 
From this perspective, QSLs are not only exotic phases of matter, but also natural laboratories in which to investigate how genuinely quantum correlations organize strongly interacting many-body systems. 
In particular, it is now well understood that long-range entanglement plays a central role in stabilizing many QSL phases and provides a unifying principle for understanding systems beyond the framework of spontaneous symmetry breaking \cite{moessner2021topological}.

At the same time, entanglement alone does not exhaust the genuinely quantum features of many-body states, nor their complexity in terms of classical representability.
For instance, highly entangled states may still admit efficient classical descriptions in special cases, while states with comparable entanglement structure can differ substantially in the resources required for their preparation, representation, or simulation. 
This observation motivates the search for complementary diagnostics capable of resolving aspects of quantum complexity that are invisible to entanglement measures alone.

The Kitaev model on the honeycomb lattice offers a particularly appealing setting in this context: it is exactly solvable, yet it realizes a highly nontrivial QSL with emergent Majorana fermions and gauge degrees of freedom \cite{kitaev2006anyons}. 
%While entanglement captures an essential aspect of the nonclassical correlations present in QSLs,
This combination of analytical tractability and intrinsically quantum many-body structure makes it an ideal platform to probe other forms of quantum complexity beyond entanglement, within the framework of quantum resource theory~\cite{chitambar2019quantum}. 
%Indeed, while entanglement captures an essential aspect of the nonclassical correlations present in QSLs, it does not by itself fully characterize the difficulty of representing or simulating a quantum state classically. 
%Quantum resource theory provides a broader language for this purpose, identifying distinct resources that can obstruct efficient classical simulation and underlie possible quantum advantage \cite{chitambar2019quantum}. 

    %Quantum spin liquids (QSLs) continue to attract intense research interest in both theoretical and experimental condensed matter physics due to their exotic quantum properties such as the absence of conventional magnetic order, fractionalized excitations, emergent gauge fields and topological order \cite{savary2016quantum,broholm2020quantum}.
    %It is now well understood that these phenomena originate from long-range entanglement which acts as the main organizing principle in many-body systems evading spontaneous symmetry breaking \cite{moessner2021topological}.
    %While this perspective suggests that entanglement plays the central role in the quantum complexity of strongly correlated matter, quantum resource theory \cite{chitambar2019quantum} identifies several aspects of quantum states that can hinder efficient classical simulation and leverage quantum advantage.

    A particularly relevant form of quantum resource is \emph{magic}, or nonstabilizerness. Originally conceived in the context of quantum error correction, where it provides a sharp distinction between easy and challenging operations, magic has gained considerable attention over the last three years in the context of many-body quantum physics, including research on quantum criticality \cite{sarkar2020characterization,white2021conformal,oliviero2022magic,tarabunga2023many,tarabunga2024critical,falcao2025nonstabilizerness,haug2025probing,leone2024phase}, non-equilibrium dynamics \cite{tirrito2025magic,wang2025magic,turkeshi2025magic,tirrito2024quantifying}, quantum chaos \cite{leone2021quantum,turkeshi2023measuring,leone2023nonstabilizerness,turkeshi2025pauli,bera2025non} and topological matter \cite{ellison2021symmetry,lami2023nonstabilizerness,nehra2025topological}.
    In short, magic quantifies the difficulty of performing classical simulations within the stabilizer formalism \cite{gottesman1997stabilizer,nielsen2010quantum} and is, in that sense, the deviation of a state from the stabilizer-state manifold \cite{bravyi2005universal,veitch2014resource}.
    The latter is a set of specific quantum states that are uniquely characterized by a set of Pauli strings and generated from computational product states using Clifford quantum circuits.
    
    In many-body systems, magic represents a notion of quantum complexity that is complementary to entanglement, as illustrated by the toric code model \cite{kitaev2003fault}.
    Namely, the toric code ground state is a paradigmatic example of a long-range entangled state with $\mathbb Z_2$ topological order but, at the same time, an exactly solvable stabilizer state with vanishing magic and thus low quantum complexity.
    Since QSLs with $\mathbb{Z}_2$ topological order have become increasingly accessible in quantum simulation experiments \cite{semeghini2021probing,satzinger2021realizing,kalinowski2023non,will2025probing,evered2025probing}, this observation naturally raises the question of the conditions under which QSLs exhibit high complexity that challenges classical simulations and efficient state preparation.
    This question becomes particularly relevant in systems close to criticality, as recent studies have shown that magic, much like entanglement, can be sensitive to phase transitions and other critical phenomena \cite{tarabunga2023many,tarabunga2024critical}.
    
    These and other works on many-body systems typically quantify magic using the stabilizer R{\'e}nyi entropy (SRE) \cite{leone2022stabilizer} of either the full wavefunction or connected subsystems \cite{tarabunga2023many,nehra2025topological}.
    The SRE is a well-defined measure of magic that, although based on an exponentially fast growing number of expectation values, can be efficiently sampled for large systems under appropriate conditions.
    Existing approaches include matrix-product-state and tensor-network techniques for sufficiently weakly entangled states \cite{haug2023quantifying,lami2023nonstabilizerness,tarabunga2024critical,lami2024unveiling,tarabunga2024nonstabilizerness} as well as Quantum Monte Carlo sampling for sign-problem-free systems \cite{ding2025evaluating}.
    More recently, a sampling algorithm was developed for \emph{fermionic Gaussian states} \cite{collura2026non} which enabled the study of magic in highly entangled free-fermion systems \cite{collura2026nonlocal} and during monitored free-fermion dynamics \cite{tirrito2025magic,wang2025magic}.
    In the context of QSLs, these results naturally raise the question of whether the latter algorithm can be adapted to the emergent fractionalized free-fermion description of the Kitaev honeycomb model (KHM).

    The KHM is an extensively studied, exactly solvable spin model on the honeycomb lattice that hosts a gapless QSL as well as gapped QSLs with $\mathbb Z_2$ and (in the presence of an effective magnetic field) non-Abelian Ising topological order \cite{kitaev2006anyons}.
    The exact solutions are facilitated by an extensive number of constants of motion, reflecting an emergent static $\mathbb Z_2$ gauge field, and a mapping to a quadratic fermionic Hamiltonian.
    This approach has been generalized and applied to other lattice geometries, including three-dimensional (3D) \cite{mandal2009exactly,eschmann2020thermodynamic,hermanns2015weyl} and completely amorphous systems \cite{cassella2023exact}, as well as to larger spin degrees of freedom \cite{wu2009gamma}.
    On the experimental side, enormous efforts have been devoted to the observation of QSLs in various spin-orbit-assisted Mott insulators that are predicted to host the characteristic exchange anisotropy of the KHM, see Refs. \cite{winter2017models,hermanns2018physics,trebst2022kitaev} for reviews.
    This research is complemented by recent analog and digital quantum simulations of the ground state or non-equilibrium dynamics of the KHM \cite{semeghini2021probing,satzinger2021realizing,kalinowski2023non,will2025probing,evered2025probing}.
    Regarding quantum complexity, various studies investigated the entanglement entropy \cite{yao2010entanglement,meichanetzidis2016anatomy ,mandal2016entanglement,dora2018gauge} and, more recently, the genuine multipartite entanglement \cite{lyu2025multiparty} of the KHM ground state.
    The former quantity comprises the topological entanglement entropy and a fermionic contribution that detects the topological phase transition between the gapless and the gapped (Abelian) phases.

    Here, we complement these studies by presenting a comprehensive study of the SRE in a two-dimensional (2D) QSL.
    We systematically improve and adapt the aforementioned algorithm for fermionic Gaussian states to sample the SRE of the KHM ground state in systems with up to $4600$ spins throughout the whole phase diagram, consisting of the gapless and $\mathbb{Z}_2$ topologically ordered phases.
    We find that SRE peaks deep in the gapless phase, while it decreases in the gapped phases, consistent with analytical estimates that we derive within perturbation theory.
    To investigate the phase transition, we compute the subleading correction to the volume law that generally governs the magic of many-body ground states \cite{ding2025evaluating}, and report that this contribution exhibits scaling behavior near the critical point, allowing for the extraction  of critical exponents.
    Our results thus establish SRE as a sensitive probe of phase transitions between distinct QSLs.
    
    Beyond criticality, volume-law corrections also provide insight into the spatial structure of many-body magic \cite{ding2025evaluating}. 
    In particular, finite corrections in the thermodynamic limit may indicate the presence of \emph{non-local magic} \cite{collura2026nonlocal, iannotti2026nonlocal}, a still poorly understood contribution that is intrinsically tied to entanglement.
    This perspective is closely related to recent conceptual works on long-range magic, which identifies the presence of magic that cannot be removed by local unitary transformations or finite-depth quantum circuits \cite{sarkar2020characterization,haug2023quantifying,ellison2021symmetry,korbany2025long,wei2026long}.
    Consistently with these studies, we find that the thermodynamic volume-law corrections vanish in the Abelian gapped phases, while they assume finite non-constant values in the gapless phase, thereby indicating a rich interplay between entanglement and magic.
    %From this perspective, our work presents a first step in the context of gapless spin liquids, and their relation to non-local magic. \tg{Don't like how this sentence has turned out}%lays the groundwork for future studies on magic in 2D or 3D long-range-entangled systems, quantum complexity of critical QSLs, and non-local magic in many-body systems.

    The remainder of this article is structured as follows. In Sec.~\ref{sec:summary}, we provide a more detailed summary of our results.
    Subsequently, in Sec.~\ref{sec:model}, we briefly review the Kitaev honeycomb model and derive the relation between arbitrary fermionic correlation functions and Pauli-string expectation values that determine the stabilizer R{\'e}nyi entropy.
    These Majorana string correlation functions can be efficiently sampled using the optimized algorithm described in Sec.~\ref{sec:sampling}.
    In Sec.~\ref{sec:num}, we present our numerical results for the SRE, including the full-state magic and the scaling analysis near the phase transitions.
    The analytical estimates from perturbation theory for the gapped phase are derived in Sec.~\ref{sec:perturbation}.
    We conclude in Sec.~\ref{sec:conclusion} and provide more technical details in the appendices.
    In App.~\ref{sec:groundstate}, we derive the correct global ground state of the KHM from Lieb's theorem for the considered system geometries and coupling constants. 
    In App.~\ref{sec:JW}, we benchmark our results by considering the decoupled limit of KHM, which is effectively described by one-dimensional (1D) spin chains.
    Finally, in App.~\ref{sec:m2}, we provide more numerical results on the SRE for R{\'e}nyi coefficient $n=2$.
    
\section{Summary of results}
    \label{sec:summary}
    \begin{figure}[t]
        \centering
        %\makebox[\textwidth][c]{\includegraphics[width=0.3\textwidth]{images/temporary_lattice.jpg}
        %\includegraphics[width=0.3\textwidth]{images/m1trsum.png}
        %\includegraphics[width=0.4\textwidth]{images/malphapert.png}}
        \includegraphics[width=\textwidth]{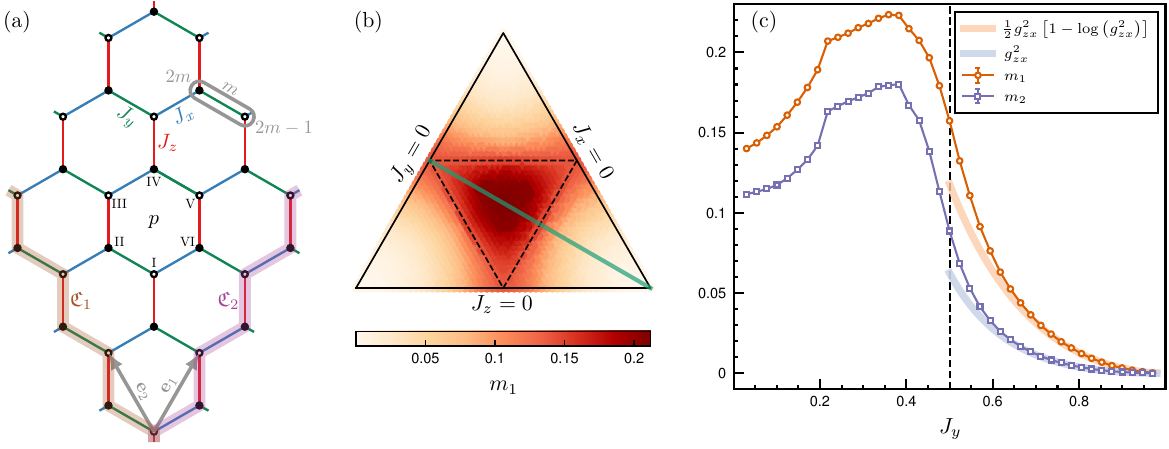}
        \caption{(a) Slice of a honeycomb lattice with the three bond types $x$, $y$ and $z$ corresponding to different edge colors. We consider lattices with periodic boundary conditions and supercell lattice vectors $L_1\mb e_1$ and $L_2\mb e_2$, where $\mb e_{1,2}=(\pm\frac{1}{2},\frac{\sqrt{3}}{2})$. (b) SRE density $m_1=M_1/N$ for $L_1=L_2=4$ within the phase diagram defined by $J_x+J_y+J_z=1$, with a gapless QSL in the center triangle and $\mathbb Z_2$ QSLs in the outer triangles. 
        (c) SRE densities $m_1$ and $m_2=M_2/N$ along the green line indicated in (b). The thick lines show analytical approximations from a perturbative theory in the coupling parameter $g_{zx}=(1-J_{y})/(4J_y)$, while symbols show numerical results for $L_1=4$ and $L_2=8$. 
        The kink around $J_y\sim0.2$ arises due to a level-crossing of flux-free sectors, see Fig.~\ref{fig:2}.
        The error bars estimated from the sample variance are significantly smaller than the used symbols.
        The used sample sizes $N_\mr s$ reach up to (a) $10^4$ and (b) $10^5$.
         }
        \label{fig:1} 
    \end{figure}
    In the following, we briefly summarize our main results.
    Throughout this work, we quantify  magic, or nonstabilizerness, using the SRE \cite{leone2022stabilizer}.
    Considering a system of $N$ spin-$1/2$ degrees of freedom, or qubits, the $n$-SRE of a pure quantum state $\rho$ is defined as
    \be
    \label{eq:sre}
        M_{n}(\rho)=\frac{1}{1-n}\log\lsb\sum_{\bs{\sigma}\in\mc P}\pi_{\rho}^{n}(\bs{\sigma})\rsb-N\log 2,
    \ee
    where the summation runs over the set of all Pauli strings 
    $\mc P=\{\sigma_1...\sigma_N\mid \sigma_j\in\{\mathbb 1_j, \sigma^x_j,\sigma^y_j,\sigma^z_j\}\}$,
    and where we have introduced the classical probability distribution of Pauli strings
    \be
    \label{eq:probsigma}
        \pi_{\rho}(\bs{\sigma})=\frac{1}{2^N}{|\Tr{\lb\bs{\sigma}\rho\rb|}^2}.
    \ee
    The $n$-SRE (for $n>0$) is an established well-defined measure of magic that can be numerically computed for many-body quantum systems, provided that an efficient sampling algorithm of the exponentially large number of Pauli strings is applicable.\footnote{The argument of the logarithm in Eq.~\eqref{eq:sre} can be rewritten as the expectation value of $\pi_{\rho}^{n-1}$, i.e., $\sum_{\bs\sigma\in\mc P}\pi_{\rho}(\bs \sigma)\pi_{\rho}^{n-1}(\bs \sigma)$. Thus, SREs can be computed by sampling strings $\bs \sigma$ with probability $\pi_{\rho}(\bs \sigma)$ and subsequently averaging $\pi_{\rho}^{n-1}(\bs \sigma)$ over these samples.}
    %In this study, we adapt and improve the recently introduced algorithm for Gaussian fermionic states \cite{collura2026non} to compute the ground-state magic of the KHM \cite{kitaev2006anyons}.
    In this study, we adapt and substantially improve the recently introduced algorithm for Gaussian fermionic states \cite{collura2026non} to compute the ground-state magic of the KHM \cite{kitaev2006anyons}.
    %While the \emph{physical} ground state of the model is highly non-Gaussian with respect to a Jordan-Wigner transformation, we employ the standard mapping of the KHM to a free-fermion Hamiltonian (with Gaussian eigenstates) by representing the spin in terms of four Majorana operators with a gauge redundancy \cite{kitaev2006anyons,feng2007topological}.
    %By deriving the correspondence between physical Pauli strings $\boldsymbol{\sigma}$ and strings of itinerant Majorana operators within a fixed gauge configuration, we establish a simple relation between the SRE and an analogous quantity in the enlarged Hilbert space.
    Although any Pauli string can, in principle, be mapped onto a string of Majorana operators, the mapping most commonly invoked in this context is the standard Jordan--Wigner transformation. 
    However, this is not the fermionization that reveals the integrable structure of the Kitaev honeycomb model: with respect to Jordan--Wigner fermions, the \emph{physical} ground state remains highly non-Gaussian, and the Gaussian-state algorithm for the SRE cannot be applied directly. 
    Here we instead exploit the genuinely different Majorana representation underlying Kitaev's exact solution, in which each spin is fractionalized into four Majorana operators subject to gauge redundancy, and the Hamiltonian becomes quadratic in the itinerant Majoranas within a fixed gauge sector \cite{kitaev2006anyons,feng2007topological}. 
    A central step of our work is to show that this nontrivial fermionization admits a precise correspondence between physical Pauli strings $\boldsymbol{\sigma}$ and strings of itinerant Majorana operators in the enlarged Hilbert space. 
    This correspondence allows us to relate the physical SRE to an analogous Majorana-resolved quantity ({\it c.f.} Eq.~(\ref{eq:srec})) computed within a fixed gauge configuration, thereby making it possible to apply the Gaussian-fermionic algorithm to the Kitaev spin liquid.
    
    Enabling the use of our optimized and highly scalable algorithm, which achieves a linear speedup in the system size with respect to the original version, reducing the computational scaling from $O(N^4)$ to $O(N^3)$, we are able to sample the SRE on lattices with up $N\simeq 4600$ spins throughout the different phases of the model.
    This allows us to investigate the thermodynamic behavior of magic in  a long-range-entangled 2D system.
    Importantly, our optimized algorithm is directly applicable to arbitrary free-fermion systems, providing a high-performance computational tool that enables the study of significantly larger system sizes than previously accessible.
    
    The standard KHM involves three coupling constants $J_x$, $J_y$ and $J_z$ associated with spin interactions along the three different bond directions on the honeycomb lattice, see Fig.~\ref{fig:1}(a).
    The model hosts a gapless quantum spin liquid (QSL) when 
    \be
        \label{eq:phase_boundary}
        J_\alpha\leq J_\beta+J_\gamma,
    \ee
    for every permutation $(\alpha,\beta,\gamma)$ of $(x,y,z)$.
    Upon introducing sufficient anisotropy, the system undergoes a transition to a gapped $\mathbb Z_2$ QSL once the inequality corresponding to a specific permutation is violated.
    %For comparable coupling constants, the KHM hosts a gapless quantum spin liquid (QSL), while upon introducing sufficient anisotropy, the system undergoes a transition to a gapped $\mathbb Z_2$ QSL at the critical lines defined by
    %\be
     %   \label{eq:phase_boundary}
     %   J_\alpha=J_\beta+J_\gamma,
    %\ee
    %where $(\alpha,\beta,\gamma)$ is a permutation of $(x,y,z)$.
    Figs.~\ref{fig:1}(b), (c) show the full-state SRE densities $m_1=M_1/N$ and $m_2=M_2/N$ throughout the triangular phase diagram of the model.
    We find that magic peaks around the isotropic point, $J_x=J_y=J_z$, while it decreases to zero in the anisotropic limit approaching a stabilizer product state of decoupled dimers.
    We perform a perturbative expansion of the Pauli strings in the latter regime to obtain analytical estimates of the SREs that quantitatively agree with our numerical results deep in the gapped phase, see Fig.~\ref{fig:1}(c).
    Notably, our perturbative analysis suggests that a similar behavior of magic in the gapped phase is expected even in the presence of the static flux excitations of the KHM.

    While the magic of ground states of local Hamiltonians is generally governed by a volume law $M_{n}\sim N$, a recent work \cite{ding2025evaluating} demonstrated that subleading volume-law corrections can be an insightful indicator of both \emph{non-local} magic and criticality.
    We fit these corrections using the assumed parametrization
    \be
    \label{eq:vollaw}
        M_{n}\simeq a_{n}N+b_{n},
    \ee
    with fitting parameters $a_{n}$ and $b_{n}$.
    We focus on the $1$-SRE and report that, in the thermodynamic limit, the correction $b_1$ vanishes throughout the gapped phase, whereas it takes finite, non-constant values in the gapless QSL. This finite contribution hints at the presence of non-local magic that cannot be removed by any local finite-depth quantum circuit \cite{korbany2025long,wei2026long,ding2025evaluating}.
    We confirm these results by computing the SRE for Hamiltonians with $J_y=0$ corresponding to the left black solid line in the phase diagram in Fig.~\ref{fig:1}(b). 
    In this limit, the system consists of decoupled 1D spin chains that can be solved using a standard Jordan-Wigner transformation, thereby enabling numerical simulations with larger linear system sizes and more precise SRE sampling.
    
    Moreover, we observe that both the SRE density $m_1\simeq a_1$  and the volume-law correction $b_1$ exhibit finite-size crossing points at the critical couplings in Eq.~\eqref{eq:phase_boundary}.
    By performing a standard finite-size scaling analysis for $m_1$ and $b_1$, we extract the critical exponents, which are consistent with the literature \cite{uryszek2020fermionic,hu2024nature}.
    Our results thus demonstrate that  magic is able to detect the topological phase transition from a gapless QSL to a $\mathbb Z_2$ topologically ordered phase, and does so according to universal critical exponents.
    
\section{Kitaev honeycomb model}
    \label{sec:model}
    
\subsection{Model Hamiltonian}
    In the following, we briefly review the Kitaev honeycomb model and its exact solution to fix our notation. 
    For a more detailed discussion, see, for example, Refs.~\cite{kitaev2006anyons,pedrocchi2011physical,zschocke2015physical}.
    The KHM \cite{kitaev2006anyons} is defined on the honeycomb lattice, which comprises three different bond types denoted by $x$, $y$ and $z$, see Fig.~\ref{fig:1}(a).
    The Hamiltonian reads
    \be
        \label{eq:hamiltonian}
        H = -J_x\sum_{\langle j,k\rangle_x}\sigma^x_j\sigma^x_k -J_y\sum_{\langle j,k\rangle_y}\sigma^y_j\sigma^y_k -J_z\sum_{\langle j,k\rangle_z}\sigma^z_j\sigma^z_k,
    \ee
    where $\sigma^\alpha_j$, with $\alpha\in\{x,y,z\}$, are Pauli operators acting on the spin-$1/2$ degree of freedom at site $j$ and $\langle j,k\rangle_\alpha$ denotes the nearest-neighbor bond of type $\alpha$ connecting the site $j$ on the odd sublattice with the site $k$ on the even sublattice.

    The KHM hosts an extensive number of constants of motion, including the plaquette operators which are associated with the hexagonal plaquettes $p$ of the lattice and are defined by
    \be
        \label{eq:plaquette}
        W_p=\sigma^z_\mr{I}\sigma^x_\mr{II}\sigma^y_\mr{III}\sigma^z_\mr{IV}\sigma^x_\mr{V}\sigma^y_\mr{VI},
    \ee
    using the convention indicated in Fig.~\ref{fig:1}(a).
    Assuming periodic boundary conditions, we further define the two conserved operators
    \be
        \label{eq:wilson}
        \mf W_1=\prod_{j\in\mathfrak C_1}\sigma_j^x,\qquad \mf W_2=\prod_{j\in\mathfrak C_2}\sigma_j^y,
    \ee
    where $\mathfrak C_1$ and $\mathfrak C_2$ are contours along inequivalent non-contractible loops winding around the system, see Fig.~\ref{fig:1}(a).
    The eigenvalues $w_p=\pm 1$ of the plaquette operators $W_p$ together with the eigenvalues $\mf w_1=\pm 1$ and $\mf w_2=\pm 1$ of $\mf W_1$ and $\mf W_2$, respectively, define decoupled eigensectors, referred to as \emph{flux sectors}.
    From the perspective of the stabilizer formalism \cite{gottesman1997stabilizer,nielsen2010quantum}, we can think of the operators $w_p W_p$, $\mf w_1\mf W_1$ and $\mf w_2\mf W_2$ as stabilizers of flux sectors.
    More specifically, for a system with $N$ spins and $N_\mr p=N/2$ plaquettes, a flux sector is stabilized by $S\in\mc S_{\mb w\bs{\mf w}}$, where $S_{\mb w\bs{\mf w}}=\braket{G_{\mb w\bs{\mf w}}}$ is the stabilizer group generated by the $N+1$ independent generators\footnote{The number of independent generators arises due to the global $\mathbb Z_2$ constraint $\prod_p W_p=\mathbb 1$.}
    \be
        \label{eq:stabilizer_flux}
        G_{\mb w\bs{\mf w}}=\{ w_1W_1,...,w_{N-1}W_{N-1},\mf w_1\mf W_1,\mf w_2\mf W_2\}.
    \ee
    In this work, we investigate the magic of the lowest-energy eigenstates within specific flux sectors.
    We obtain the exact expressions for these eigenstates by means of the spin representation $\sigma^\alpha_j=ib_j^\alpha c_j$, where $c_j$, $b_j^x$, $b_j^y$ and $b_j^z$ are Majorana operators with standard Majorana anticommutation algebra \cite{kitaev2006anyons}.
    Since the employed representation artificially enlarges the local Hilbert space dimension, we further impose the local constraints
    \be
        \label{eq:constraint}
        D_j\ket{\Psi}=\ket{\Psi},\quad D_j=b_j^xb_j^yb_y^zc_j,
    \ee
    on physical states $\ket{\Psi}$ in the original Hilbert space.
    The key step in the solution of the KHM is the identification of the mutually commuting operators $\hat{u}_{jk}=ib^\alpha_jb^\alpha_k$ for every bond $\braket{j,k}_\alpha$, which reflect an emerging $\mathbb Z_2$ gauge field.
    We can replace the operators $\hat{u}_{jk}$ with their appropriately chosen eigenvalues $u_{jk}=\pm 1$ to define a gauge configuration $\mb u=\{u_{jk}\}$ for the selected flux sector.
    Upon this substitution, we obtain a quadratic fermionic Hamiltonian $H_\mb{u}=-i\sum_{\braket{jk}_\alpha}J_\alpha u_{jk}c_jc_k$.
    By pairing itinerant Majoranas connected by $y$ bonds, see Fig.~\ref{fig:1}(a), we define the complex fermionic annihilation and creation operators
    \be
        \label{eq:fermions}
        f_m=\frac{1}{2}\lb c_{2m-1}+ic_{2m}\rb,\qquad f_m^\dagger=\frac{1}{2}\lb c_{2m-1}-ic_{2m}\rb,
    \ee
    respectively.
    We may then rewrite the quadratic Hamiltonian for a specified gauge configuration $\mb u$ as
    \be
        \label{eq:H_u}
        H_\mb{u}=\frac{1}{2}\mb f^\dagger M_\mb u\mb f,\qquad \mb f=\bmat f\\ f^\dagger\emat,
    \ee
    where $f$ ($f^\dagger$) is a column vector composed of the complex fermionic annihilation (creation) operators and where the $N\times N$ matrix $M_\mb u$ is of the form
    \be
        M_\mb u=\bmat A_\mb u & B_\mb u\\ -B_\mb u^* & -A_\mb u^*\emat,
    \ee
    with the $N_\mr p\times N_\mr p$ matrices $A_\mb u=A_\mb u^\dagger$ and $B_\mb u=-B_\mb u^T$.
    The Hamiltonian is diagonalized by a standard Bogoliubov transformation \cite{blaizot1986quantum},
    \be
        \label{eq:bogoliubov}
        \bmat f \\ f^\dagger\emat=Q_\mb u\bmat a\\ a^\dagger \emat, 
    \ee
    where the unitary matrix $Q_\mb u$ is chosen such that $Q_\mb u^\dagger M_\mb uQ_\mb u=\mr{diag}\,(\varepsilon_1,...,\varepsilon_{N_\mr p}, -\varepsilon_1,...,-\varepsilon_{N_\mr p})$ for the single-particle energies $0\leq \varepsilon_1\leq ...\leq\varepsilon_{N_\mr p}$ and, as before, $a$ ($a^\dagger$) is a column vector composed of the complex fermionic annihilation (creation) operators $a_\nu$ ($a_\nu^\dagger$).
    Inserting the transformation~\eqref{eq:bogoliubov} into Eq.~\eqref{eq:H_u} yields the diagonal Hamiltonian
    \be
    \label{eq:diagham}
        H_\mb u=\sum_{\nu=1}^{N_\mr p} \varepsilon_\nu\lb a^\dagger_\nu a_\nu-\frac{1}{2}\rb.
    \ee
    While this concludes the solution for the fermionic spectrum of a specified gauge configuration in the enlarged Hilbert space, we note that the constraints on physical states in Eq.~\eqref{eq:constraint}  select a fermionic parity $(-1)^{N_a}$, with $N_a=\sum_\nu a_\nu^\dagger a_\nu$, for the fermionic eigenmodes.
    Indeed, the correct fermionic parity can be extracted from the projector onto the physical Hilbert space \cite{kitaev2006anyons,pedrocchi2011physical,zschocke2015physical}
    \be
    \label{eq:pphys}
        \Pi_\mr{phys}=\prod_j\frac{1+D_j}{2}=\frac{1}{2}S_{\mr{phys}}\lsb 1+(-1)^{N_a+\theta}\det{(Q_\mb u)}\prod_{\braket{j,k}_\alpha}u_{jk}\rsb,
    \ee
    where $S_{\mr{phys}}$ is an operator that symmetrizes over gauge-equivalent configurations,\footnote{The exact expression of $S_{\mr{phys}}$ is not relevant for the computation of the SRE~\eqref{eq:sre} as it is based on gauge invariant spin correlation functions which assume the same value in every gauge-equivalent configuration \cite{baskaran2007exact}.}% as it is based on correlation functions of the physical spin degrees of freedom.} 
    see Ref.~\cite{pedrocchi2011physical}, and $\theta$ is an integer that generally depends on the system geometry and boundary conditions, see App.~\ref{sec:groundstate} for details.
    
    We carefully employ Lieb's theorem \cite{lieb1994flux,kitaev2006anyons} to ensure that the ground state of the KHM corresponds to the fermionic vacuum in a \emph{vortex-free} sector, i.e., with $W_p=1$ for all plaquettes $p$, provided that the physical ground state has even parity.
    To this end, we minimize the energy over the four vortex-free sectors distinguished by the two Wilson loops in Eq.~\eqref{eq:wilson} and check the fermionic parity of the global ground state.
    This procedure eliminates an established effect which is limited to small systems in the gapless phase but may persist up to arbitrarily large sizes in the gapped phase \cite{pedrocchi2011physical,zschocke2015physical}.
    Interestingly, since the splitting of the topological ground-state degeneracy generally depends on the system size and geometry, this routine gives rise to kinks in the system-size dependence of the computed quantities for the global ground state, see Sec.~\ref{sec:num} and App.~\ref{sec:groundstate}, whose origin we are in full control of.
    We conclude our review of the KHM by noting that the vortex-free sector with the translational invariant gauge configuration $\mb u=\{u_{jk}=1\}$ allows for a diagonalization in momentum space that yields the phase boundaries defined by Eq.~\eqref{eq:phase_boundary}. 
    More specifically, the fermionic spectrum is gapped whenever $J_\alpha > J_\beta+J_\gamma$ for a permutation $(\alpha, \beta, \gamma)$ of $(x,y,z)$ and gapless for sufficiently isotropic coupling constants, i.e., if $J_\alpha \leq J_\beta+J_\gamma$ for each permutation \cite{kitaev2006anyons}.

\subsection{Evaluation of Pauli strings}

Above we solved the KHM by mapping the model to a free-fermion Hamiltonian within a fixed gauge configuration.
Free-fermion Hamiltonians are diagonalized by fermionic Gaussian states, which are fully characterized by their covariance matrix~\cite{surace2022fermionic}.
In our case, the covariance matrix reads
\begin{equation}
\label{eq:gamma}
    \Gamma_{mn}(\rho) = -i\Big(2\Omega^*C(\rho)\Omega^T-1\Big)_{mn},
\end{equation}
where $C_{mn}(\rho) = \mathrm{Tr}(\mb{f}^{\dagger}_{m}\mb{f}_{n})$ is the standard fermionic correlation matrix, $\mb{f}_{m}=f_m$ and $\mb{f}_{m+N_\mr p}=f^{\dagger}_m$ are the complex fermionic operators defined in Eq.~\eqref{eq:fermions} and $\Omega$ is the unitary transformation of complex Dirac fermions to Majorana operators, see Ref.~\cite{surace2022fermionic}.
Recently, Ref.~\cite{collura2026non} introduced an algorithm that, given the covariance matrix of a fermionic Gaussian state, efficiently samples all fermionic correlation functions or, more specifically, the expectation values of all Majorana strings $\mb c\in\mc C$, where
\begin{equation}
    \label{eq:cstrings}
    \mc C = \{c_1^{x_1}...c_N^{x_N} \mid \ x_j\in\{0,1\}\}
\end{equation}
is the set of all Majorana strings for a system with $N$ different Majorana operators.
{Here, $x_j\in\{0,1\}$ is a binary index specifying whether the Majorana operator $c_j$ is absent ($x_j=0$) or present ($x_j=1$) in a given string.}

We describe an optimized version of this algorithm in Sec.~\ref{sec:sampling}.
The underlying idea is to sample Majorana strings $\mb c$ from the probability distribution 
\be
    \label{eq:probc}
    \pi_{\rho}(\mb{c}) = \frac{|\mathrm{Tr}[\mb{c}\rho]|^2}{2^{N_\mr p}},
\ee
where $\rho$ is the pure state of interest and $N_\mr p=N/2$.
Notably, $\pi_{\rho}(\mb{c})$ is perfectly equivalent to the probability distribution of Pauli strings obtained from a standard Jordan-Wigner transformation, which represents one Pauli operator in terms of two Majorana operators \cite{collura2026non}.
The SRE~\eqref{eq:sre} of these $N_\mr p$ spin systems can then be written as
\be
    \label{eq:srec}
        M_{n}^{c}(\rho)=\frac{1}{1-n}\log\lsb\sum_{\mb{c}\in\mc C}\pi_{\rho}^{n}(\mb{c})\rsb-N_\mr p\log 2.
\ee
In our case, however, we employed an overcomplete Majorana representation of the $N$ spins of the KHM to obtain a fermionic Gaussian eigenstate defined by $N$ itinerant Majorana operators $c_j$ within a fixed gauge configuration.
It is therefore not obvious how the formal equivalent of the SRE in Eq.~\eqref{eq:srec} is related to the actual SRE of the KHM ground state $\rho$, and whether the algorithm in Ref.~\cite{collura2026non} can be applied.
In what follows, we show that
\begin{equation}
\label{eq:mageq}
    M_{n}^c(\rho)= M_{n}(\rho)
\end{equation}
actually holds and thereby ensure that the results obtained from the Majorana sampling technique indeed measure the magic of the state $\rho$.

The derivation of Eq.~\eqref{eq:mageq} is based on the established result that the spin-spin correlations $\Tr{\lsb\rho \sigma^\alpha_j\sigma^\beta_k\rsb}$ of the KHM are ultra-local \cite{baskaran2007exact} and, more specifically, only finite if $\alpha=\beta$ and if $j$ and $k$ are connected by the nearest-neighbor bond $\braket{j,k}_\alpha$. %\tg{By combining this statement with Wick's theorem, we understand that}
\emph{Relevant} Pauli strings (i.e., with finite expectation value) are then obtained by multiplying bond operators $\sigma^\alpha_j\sigma^\alpha_k$ for bonds $\braket{j,k}_\alpha$,\footnote{This is because these bond operators only generate Pauli strings that commute with the constants of motion in Eq.~\eqref{eq:stabilizer_flux} and thus do not couple different flux sectors.} and are conveniently parametrized within the framework of graph theory.
To this end, we may describe the honeycomb lattice as a directed three-edge-colored graph $\mc G=(\mc V, \mc E)$ with vertices $\mc V=\{1,...,N\}$ and edges
\be
   \mc E=\{(j,k;\alpha)\}.
\ee
The triplets $(j,k;\alpha)\in\mc E$ include the vertices $j$ and $k$ connected by the edge of color $\alpha\in\{x,y,z\}$ corresponding to the bond $\braket{j,k}_\alpha$.
A given subset $\mc X\subseteq \mc E$ defines a relevant Pauli string $\bs\sigma_{\mc X}\in\mc P$ in terms of the parametrization
    \be
    \label{eq:relevantsigma}
        \bs \sigma_{\mc X}=\mathrm{e}^{-i\theta_{\mc X}}\prod_{(j,k;\alpha)\in\mc X}\sigma^\alpha_j\sigma^\alpha_k,
    \ee
    where $\mathrm{e}^{-i\theta_{\mc X}}$ compensates for any global complex phase.
Within the Majorana representation $\sigma^\alpha_j\sigma^\alpha_k=-i\hat{u}_{jk}c_jc_k$, we may then select a gauge configuration $\mb u=\{u_{jk}\}$ of the flux sector of interest, by replacing the operators $\hat{u}_{jk}$ with their eigenvalues $u_{jk}$, to obtain the corresponding Majorana string
\be
    \mb c_{\mc X}=\mathrm{e}^{-i\phi_{\mc X}}\prod_{(j,k;\alpha)\in\mc X} c_jc_k\in\mc C,
\ee
where, again, $\mathrm{e}^{-i\phi_{\mc X}}$ compensates for any global complex phase, including the sign that may arise when ordering the Majorana operators in increasing order, see Eq.~\eqref{eq:cstrings}.

\paragraph*{Relation between Pauli and Majorana strings of complementary subgraphs.}
The first key observation for the the proof of Eq.~\eqref{eq:mageq} is that there are exactly two subsets $\mc X_1,\mc X_2\subseteq \mc E$ that map to the same relevant Pauli string.
    In particular, these subsets  $\mc X_1$ and $\mc X_2$ are complementary with respect to each other, i.e., $\mc X_2=\bar{\mc X_1}=\mc E\setminus\mc X_1$, such that
    \be
        \label{eq:paulistrings_complement}
        \bs \sigma_{\mc X}=\bs \sigma_{\bar{\mc X}}
    \ee
    for every $\mc X\subseteq \mc E$.
    On the other hand, however, the two subsets $\mc X$ and $\bar{\mc X}$ map to two different Majorana strings $\mb c_{\mc X}\neq \mb c_{\bar{\mc X}}$, which are related by
    \be
        \mb c_{\mc X}=D\mb c_{\bar{\mc X}},
    \ee
    where $D=\prod_{j\in\mc V} D_j$ is the global Majorana parity.\footnote{Note that every physical state $\ket{\Psi}$ satisfies $D\ket{\Psi}=\ket{\Psi}$, see Eq.~\eqref{eq:constraint}.}
    
    To see this, notice that, since $\mc G$ is tricoordinated, each site $j\in\mc V$ appears three times in $\mathcal{E}$, once for each bond type $\alpha$. 
    For a given bipartition $\mc X\cup \bar{\mc X}=\mc E$, we may then distinguish two cases for every site $j\in\mc V$. Either (i) all bonds involving $j$ belong to the same subset, say $\mc X$, or (ii) exactly two bonds involving $j$ appear in the same subset, say $\mc X$, while the remaining one belongs to the other subset $\bar{\mc X}$.
    We then find that:
\begin{enumerate}[label=(\roman*)]
    \item site $j$ merely contributes (up to a complex phase) $\mathbb 1_j$ to $\bs \sigma_{\mc X}$ and $\bs \sigma_{\bar{\mc X}}$, while $c_j$ is included in $\mb c_{\mc X}$ but not in $\mb c_{\bar{\mc X}}$,
    \item site $j$ contributes (up to a complex phase) the same Pauli operator to $\bs\sigma_{\mc X}$ and $\bs\sigma_{\bar{\mc X}}$, while $c_j$ is included in $\mb c_{\bar{\mc X}}$ but not in $\mb c_{\mc X}$.
\end{enumerate}
Within the framework of graph theory, we can summarize these two cases by considering the spanning subgraph $\mc G_\mc X=(\mc V,\mc X)$ for $\mc X\subseteq\mc E$.
Using $c_j^2=1$, we infer that the Majorana string $\mb c_\mc X$ only includes Majorana operators on vertices with odd coordination number with respect to $\mc G_\mc X$.
On the other hand, the Pauli string $\bs\sigma_\mc X$ only involves Pauli operators on vertices with coordination number $1$ or $2$, and the same Pauli operators are included in $\bs\sigma_{\bar{\mc X}}$ by virtue of the Pauli algebra.

\paragraph*{Majorana strings and loop configurations.} The second important observation is that there are $2^{N_\mr p+1}$ different subsets ${\mc{X}}\subseteq\mc E$ that map to the same Majorana string $\mb c_{\mc X}$.
This is due to the fact that the KHM hosts $N_\mr p+1$ independent constants of motion, see Eq.~\eqref{eq:stabilizer_flux}, which merely contribute to the global phase when mapped to the corresponding Majorana string.
In other words, we find that
\be
\label{eq:majorana_string_const}
\mb c_{\mc X} = \mb c_{(\mc X \cup \mc W)\setminus (\mc X \cap \mc W)}
\ee
for any subset $\mc W\subseteq\mc E$ that defines a constant of motion $\bs 
\sigma_\mc W$ by virtue of Eq.~\eqref{eq:relevantsigma}.\footnote{Within graph theory, this result stems from the fact that constants of motion correspond to Wilson loops and that adding loops in a spanning subgraph $\mc G_\mc X$ does not change the parity of coordination numbers of vertices.}

\paragraph*{Proof of Eq.~\eqref{eq:mageq}.}
Let us combine the above results to prove Eq.~\eqref{eq:mageq} by considering the summation over all Pauli strings in the SRE~\eqref{eq:sre}.
We first restrict the summation to the relevant Pauli strings~\eqref{eq:relevantsigma} parametrized by subsets $\mc X \subseteq \mc E$, where an additional factor $\frac{1}{2}$ accounts for the fact that each $\bs \sigma_{\mc X} \in \mc P$ is generated by two subsets, see Eq.~\eqref{eq:paulistrings_complement}.
Upon substituting $\bs \sigma_{\mc X}$ by $\mb c_{\mc X}$, we finally employ Eq.~\eqref{eq:majorana_string_const} to switch to the summation over Majorana strings $\mb c \in \mc C$.
We obtain
    \be
        \sum_{\bs \sigma\in \mc P}\Tr{\lb\rho\bs \sigma\rb}^{2n}=\frac{1}{2}\sum_{\mc X\subseteq \mc E}\Tr{\lb\rho\bs\sigma_{\mc X}\rb}^{2n}=\frac{1}{2}\sum_{\mc X\subseteq \mc E}\Tr{\lb\rho\mb c_{\mc X}\rb}^{2n}=2^{N_{\mr p}}\sum_{\mb c\in\mc C}\Tr{\lb\rho\mb c\rb}^{2n}.
    \ee
Inserting this result in Eq.~\eqref{eq:srec} accounts for the correct shift of the logarithm, and concludes our derivation of Eq.~\eqref{eq:mageq}.

A similar reasoning, which we reserve for a future publication, was carried out to demonstrate that the result in Eq.~\eqref{eq:mageq} also applies to the SRE of a subsystem of the KHM, which is needed to calculate mutual magic and magic-based topological invariants \cite{tarabunga2023many,nehra2025topological}. 
Moreover, notice that the only necessary ingredients in our derivation are Pauli and Majorana operators defined on a three-edge-colored graph.
This ensures the validity of Eq.~\eqref{eq:mageq} in more general contexts such as systems with bond disorder or Kitaev interactions defined on other three-edged-colored graphs, see for example Refs. \cite{mandal2009exactly,eschmann2020thermodynamic,hermanns2015weyl}.
In fact, we numerically confirmed Eq.~\eqref{eq:mageq} for small small sizes and for both clean and bond-disordered systems up to machine precision.

We further note that, while a solution of the Kitaev honeycomb model in terms of a Jordan–Wigner transformation is available, the procedure involves the decoupling of a quartic term, which again introduces a gauge redundancy \cite{feng2007topological}. It is therefore not obvious whether or how this or other solution schemes \cite{vidal2008perturbative,kells2009description} can enable an efficient sampling routine for the SRE~\eqref{eq:sre}.

\section{Sampling algorithm}
\label{sec:sampling}
 The numerical results in this work are obtained using an optimized version of the algorithm presented in Ref.~\cite{collura2026non}, which we detail in the following.
Given the $N\times N$ covariance matrix $\Gamma$ of a Gaussian state $\rho$ with matrix elements defined by Eq.~\eqref{eq:gamma}, this algorithm efficiently samples the probability distribution~\eqref{eq:probc} of Majorana strings $\mb c$.
    In what follows, it is convenient to parametrize Majorana strings $\mb c_{\bs x}$ in terms of bit strings $\bs x\in\{0, 1\}^N$ using
    \be
        \mb c_{\bs x}=c_1^{x_1}...c^{x_N}_N.
    \ee
To simplify our notation, we will also write the probability distribution of Majorana strings as $\pi_{\rho}(\bs x)$.
The key idea of the algorithm is to deal with the exponentially large number of Majorana strings, $2^N$, by sequentially sampling $\pi_{\rho}(\bs x)$ written in terms of conditional and marginal probabilities:
\be
\label{eq:probtotal}
        \pi_{\rho}(\bs{x}) = \pi_{\rho}(x_1)\pi_{\rho}(x_2|x_1)...\pi_{\rho}(x_{N}|x_1...x_{N-1}),
    \ee
where $\pi_{\rho}(x_{\mu}|...x_{\mu-1}) = \frac{\pi_{\rho}(x_1...x_{\mu})}{\pi_{\rho}(x_1...x_{\mu-1})}$ denotes the conditional probability that bit $\mu$ equals $x_\mu$, given the string $x_1...x_{\mu-1}$.
    We can compute this quantity using the identity \cite{launay2020exact,bianchi2021page,collura2026non}
    \be
    \label{eq:pirhomu}
        \pi_{\rho}(x_1...x_{\mu}) = \frac{\det [(\mathbb{1}_{[\mu+1,N]}+\Gamma)|_{(x_1,...,x_{\mu},1,...,1)}]}{\det [\mathbb{1}+\Gamma]},
    \ee
    where $\mathbb{1}_{[\mu+1,N]}$ is the diagonal matrix with entries equal to $1$ in the interval $[\mu+1,N]$ and $0$ otherwise, and the notation $|_{(x_1,...,x_N)}$ indicates that we are restricting to the rows and columns whose indices $i$ are associated to $x_i=1$.
    The algorithm presented in Ref.~\cite{collura2026non} is an iterative process in which, at each step $\mu$, the value of $x_\mu$ is sampled with probability $\pi_{\rho}(x_{\mu}|x_1...x_{\mu-1})$ and the total probability in Eq.~\eqref{eq:probtotal} is updated accordingly.
    After the final step, $\mu=N$, the algorithm returns a bit string $\bs x$ and the probability $\pi_\rho(\bs x)$.
    
    Notably, since this procedure is based on computing the determinants of $N$ submatrices of $\Gamma$, with edge dimensions ranging from $1$ to $N$, the computational cost scales like $\mc O\lb N^4\rb$.
    In our optimized routine, we use simple algebraic identities to avoid the need of computing the determinants from scratch after every iteration, and thereby achieve a substantial speedup.
    To this end, our algorithm takes as inputs $A^{-1}=(\mathbb{I}+\Gamma)^{-1}$ and $\det{A}$ and updates these quantities after each step using the following routine.
{For each step $\mu$ we compute $\pi_{\rho}(x_1...x_{\mu-1}0)$ depending on the value of the bit $x_{\mu-1}$:
\begin{enumerate}[label=(\roman*)]
    \item For $x_{\mu-1}=0$, Eq.~\eqref{eq:pirhomu} gives
    \begin{equation}
    \begin{split}
    \det[\mathbb{1}+\Gamma]\pi_{\rho}(x_1...x_{\mu-2}00) &= \det[(\mathbb{1}_{[\mu+1,N]}+\Gamma)|_{(x_1,...,x_{\mu-2},0,0,1,...,1)}]\\ &= \det[(\mathbb{1}_{[\mu,N]}+\Gamma)|_{(x_1,...,x_{\mu-2},0,0,1,...,1)}],
    \end{split}
\end{equation}
which amounts to calculating the determinant of the matrix A \emph{computed} at step $\mu-1$ (i.e., the matrix whose determinant equals $\det[\mathbb{1}+\Gamma]\pi_{\rho}(x_1...x_{\mu-2}0) $) after removing its column and row $\mu$. 
This can be readily done with a Schur complement inverse downgrade \cite{schurscomplement}: up to a permutation that moves $\beta$ to the last position, given $A^{-1}=\begin{pmatrix}
    B & w\\
    z^T &\beta
\end{pmatrix}$, removing the last row and column implies
\begin{equation}
\label{schur}
A^{-1} \rightarrow B-\frac{w z^T}{\beta}, \quad \det A \rightarrow \beta\det A.
\end{equation}
\item For $x_{\mu-1}=1$, we obtain
\begin{equation}
    \begin{split}
    \det [\mathbb{1}+\Gamma]\pi_{\rho}(x_1...x_{\mu-2}10) &= \det [(\mathbb{1}_{[\mu+1,N]}+\Gamma)|_{(x_1,...,x_{\mu-2},1,0,1,...,1)}]\\ 
    &= \det [(\mathbb{1}_{[\mu,N]}+\Gamma)|_{(x_1,...,x_{\mu-2},1,0,1,...,1)}]\\
    &= \det [(\mathbb{1}_{[\mu-1,N]}+\Gamma-e_{\mu-1}e_{\mu-1}^T)|_{(x_1,...,x_{\mu-2},1,0,1,...,1)}],
    \end{split}
\end{equation}
which is the determinant of the matrix A \emph{associated to} step $\mu-2$ (i.e., the matrix whose determinant equals $\det[\mathbb{1}+\Gamma]\pi_{\rho}(x_1...x_{\mu-2}) $) with $1$ subtracted from the diagonal at position $\mu-1$ and the column $\mu$ and the row $\mu$ removed.
The second operation is again a Schur complement inverse downgrade, while the first one can be performed using Sherman-Morrison formula \cite{shermanmorrison}:
\begin{equation}
    \label{shermor}
    (A-e_{\mu-1}e_{\mu-1}^T)^{-1} = A^{-1} + \frac{A^{-1}e_{\mu-1}e_{\mu-1}^TA^{-1}}{1-A^{-1}_{\mu-1,\mu-1}}, \quad \det [A-e_{\mu-1}e_{\mu-1}^T] = (1-A^{-1}_{\mu-1\mu-1})\det A.
\end{equation}
\end{enumerate}
Notably, the code only stores a single matrix $A^{-1}$ and value of $\det A$, which are updated at each step. Indeed, at step $\mu$ the algorithm only needs $A^{-1}$ and $\det A$ evaluated at step $\mu-1$ if $x_{\mu-1}=0$, while it only needs the ones associated to step $\mu-2$ if $x_{\mu-1}=1$ (i.e., obtained after the Schur's complement performed at step $\mu-2$ if $x_{\mu-2}=0$ or after employing the Sherman-Morrison formula at step $\mu-1$ if $x_{\mu-2}=1$). This is equivalent to saying that at step $\mu$ we first apply Sherman-Morrison if $x_{\mu-1}=1$. Then, we update the inverse and the determinant with the results of Schur's complement performed at step $\mu$ if $x_{\mu}=0$ gets extracted, discarding the old values, while we do the opposite if $x_{\mu}=1$.\\
For what concerns the computational cost, at a step in which $A^{-1}$ has edge dimension $k$,
both the Sherman–Morrison update and the Schur-complement downgrade have a cost of $\sim k^2$, implying a total cost $\sim\mc O\lb N^3\rb$ for the whole algorithm. The required computational resources are thus reduced by a factor $N$ with respect to the version in Ref.~\cite{collura2026non} for the worst-case configuration, i.e., $x_{\mu}=1$ for all bits $\mu$.
We have verified that the actual speedup for a generic sample is $\sim\mc O\lb N^{0.6}\rb$. We emphasize that, without such a speed-up, the calculations for the volumes accessed below would have been prohibitively time-consuming.
As a last observation, notice that in general the dimension of $A^{-1}$ changes at each step according to the results of the extractions.
This must be taken care of by mapping $\mu$ into the corresponding effective index while performing operations Eqs.~\eqref{schur} and \eqref{shermor}, using $\tilde{\mu} = \sum_{i<\mu}x_i$.
We finally report our algorithm written in the logarithmic basis:

\begin{algorithm}[H]
\caption{Improved Majorana sampling}\label{alg:QA}
\begin{flushleft}
\hspace*{\algorithmicindent} \textbf{Input}: The $N \times N$ covariance matrix $\Gamma$ of the Gaussian state
\end{flushleft}
\begin{algorithmic}[1]
\State Compute $(\mathbb{I}+\Gamma)^{-1}$ and $\log\det [\mathbb{I}+\Gamma]~(= N/2\log 2$ for a pure state)
\State Initialize $\bs{x}=()$, $\log\Pi = 0$, $\log\det A=\log\det [\mathbb{I}+\Gamma]$, $A^{-1}=(\mathbb{I}+\Gamma)^{-1}$
\For{($\mu=0$, $\mu=N-1$, $\mu++$)}
\If{$\mu>0$, $x_{\mu-1}=1$} 
\State $\gamma = 1 - A^{-1}[\tilde{\mu}-1,\tilde{\mu}-1] $   
\State $\log\det A \ \texttt{+=} \ \log\gamma$
\State  $A^{-1} \ \texttt{+=} \ A^{-1}[:,\tilde{\mu}-1]\cdot A^{-1}[\tilde{\mu}-1, :] \ /\ \gamma   $

\EndIf
     \State Compute 
    $\log\pi_{\rho}(0|\bs{x}) = \log\det A + \log A^{-1}[\tilde{\mu},\tilde{\mu}] - \log\det [\mathbb{I}+\Gamma]-\log\Pi$
     \State  Set $x_{\mu}$ to 0 or 1 randomly with the logarithm of the probability being $\log\pi_{\rho}(0|\bs{x})$ or \indent $\log(1-\exp\log\pi_{\rho}(0|\bs{x}))$
     \State Update $\log\Pi\rightarrow\log\Pi+\log\pi_{\rho}(x_{\mu}|\bs{x})$, $\bs{x}\rightarrow (\bs{x},x_{\mu})$
\If{$x_{\mu}=0$}
\State $\mathrm{idx} = \mathrm{range}(A^{-1}.\mathrm{shape}[0]) \neq \tilde{\mu}$
    \State $\ A^{-1} = A^{-1}[\mathrm{idx}, \mathrm{idx}] - A^{-1}[\mathrm{idx}, \tilde{\mu}]  \cdot A^{-1}[\tilde{\mu}, \mathrm{idx}]\ / \ A^{-1}[\tilde{\mu},\tilde{\mu}]$\State $ \log\det A \ \texttt{+=} \ \log A^{-1}[\tilde{\mu},\tilde{\mu}]$
\EndIf
\EndFor
\end{algorithmic}
\begin{flushleft}
\hspace*{\algorithmicindent} \textbf{Output}: A string $\bs x \in \{0,1\}^{N}$ and the probability $\pi_{\rho}(\bs x)$
\end{flushleft}
\end{algorithm}

\section{Numerical results}
    \begin{figure}[t]
        \centering
        %\makebox[\textwidth][c]{\includegraphics[width=0.5\textwidth]{images/m12d.png}\hfill
        %\includegraphics[width=0.24\textwidth]{images/m1cfr.png}
        %\includegraphics[width=0.26\textwidth]{images/e1e2gap.png}}
        \includegraphics[width=\textwidth]{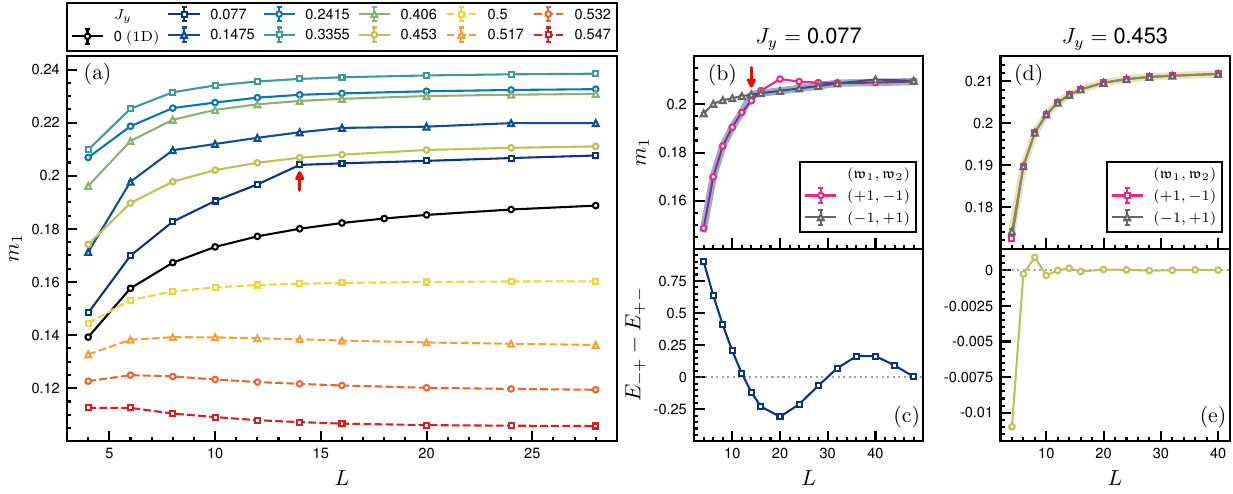}
        \caption{
        (a) Ground-state SRE density $m_1$ vs linear system size $L$ for different values of $J_y=1-2J_x=1-2J_z$ along the green line in Fig.~\ref{fig:1}(b), with dashed (solid) lines corresponding to the gapped (gapless) phase.
        The data for $J_y>0$ were obtained for a 2D system with linear system sizes $L_1=L_2=L$ and twisted periodic boundary conditions, with sample sizes $N_\mr s$ ranging from $10^4$ for $J_y\sim0$ to $10^5$ for $J_y\sim 0.5$.
        The data for the decoupled limit $J_y=0$ were obtained for a 1D spin chain and $N_\mr s=10^7$, see App.~\ref{sec:JW}.
        (b), (d) $m_1$ vs $L$ for vacuum states in different vortex-free sectors parametrized by $(\mf w_1, \mf w_2)$ at (b) $J_y=0.077$ and (d) $J_y=0.453$. The corresponding energy differences $E_{+-}-E_{-+}$ are shown in (c) and (e), respectively.
        The energy-level crossing in (c) at $L=14$ explains the kink of the ground-state SRE (thick solid lines) indicated by the red arrows in (a) and (b).
        The remaining kinks observed at low $J_y$ and in Fig.~\ref{fig:1}(c) are like-wise explained by energy-level crossings.}
        \label{fig:2}
    \end{figure}
    \label{sec:num}
    In this section, we report our numerical results for the SRE of the KHM ground state as obtained from the sampling algorithm presented in Sec.~\ref{sec:sampling} with at least $N_\mr s=10^4$ samples.
    We study two different scalable system geometries that, by virtue of Lieb's theorem \cite{lieb1994flux}, ensure a vortex-free ground state.
    While the results in Fig.~\ref{fig:1}(c) are obtained for a lattice with linear sizes $L_1=L$ and $L_2=2L$ and conventional periodic boundary conditions, all following results were computed for equal linear sizes $L_1=L_2=L$ and twisted periodic boundary conditions \cite{pedrocchi2011physical,zschocke2015physical}, see App.~\ref{sec:groundstate} for details.
    In all cases, we identify the ground state by minimizing over the four vortex-free configurations parametrized by the eigenvalues $(\mf w_1, \mf w_2)$ of the non-contractible Wilson loops~\eqref{eq:wilson}, see Sec.~\ref{sec:model}.

\subsection{Full-state stabilizer R\'eny entropy}
    For our discussion, we focus on the green line through the phase diagram in Fig.~\ref{fig:1}(b), which is parametrized by the coupling constant $J_y = 1-2J_x = 1-2J_z$, as it is covered by Lieb's theorem and includes the highly symmetric point $J_x=J_y=J_z$.
    Along this line, the system exhibits a phase transition between the gapless and gapped phase at $J_y=0.5$, while it approaches the limit of decoupled 1D spin chains for $J_y\rightarrow0$, see App.~\ref{sec:JW}, and the limit of decoupled spin dimers for $J_y\rightarrow 1$, see Sec.~\ref{sec:perturbation}.

    In Fig.~\ref{fig:2}(a), we show the system-size dependence of the $1$-SRE density $m_1=M_1/N$, where $N=2L^2$ is the number of spins.
    We find that $m_1$ approaches its thermodynamic values with positive (negative) curvature in the gapless (gapped) phase, $J_y<0.5$ ($J_y>0.5$), but exhibits various kinks as a function of $L$, e.g. for $J_y=0.077$ at $L=14$.
    These kinks are readily explained by level crossings of the vacuum energies of the different vortex-free sectors as functions of $L$, see Figs.~\ref{fig:2}(b)-(e).
    While these energies are expected to become four-fold degenerate in the thermodynamic limit, the degeneracy can be fully or partially lifted for finite system sizes.
    For example for $J_y=0.077$, the eigenvalues $(\mf w_1, \mf w_2)$ of the global ground state change from ${(+1, -1)}$ to $(-1, +1)$ when the linear system size is increased from $L=12$ to $L=14$ and give rise to the aforementioned kink, see Fig.~\ref{fig:2}(b),(c).
    For increasing $J_y$, however, these oscillations of the energy gap of different vortex-free sectors become negligible and result in a smooth convergence of $m_1$, see for example the behavior for $J_y=0.453$ in Fig.~\ref{fig:2}(d),(e).
    We perform a similar analysis of finite-size effects of the $2$-SRE $M_2$ in App.~\ref{sec:m2} and obtain results consistent with the different geometry studied in Fig.~\ref{fig:1}(c).
    Namely, we find that both densities $m_1$ and $m_2=M_2/N$ attain their maximum near the isotropic point, $J_y=\frac{1}{3}$, decrease to a finite value upon approaching the limit of decoupled 1D spin chains, $J_y=0$, and vanish in the limit of decoupled dimers, $J_y=1$.
    The results in the latter regime are consistent with analytical estimates obtained form perturbation theory, see Sec.~\ref{sec:perturbation}.

\subsection{Volume-law correction and critical behaviour}
    \begin{figure}[t]
        \centering
        \includegraphics[width=\textwidth]{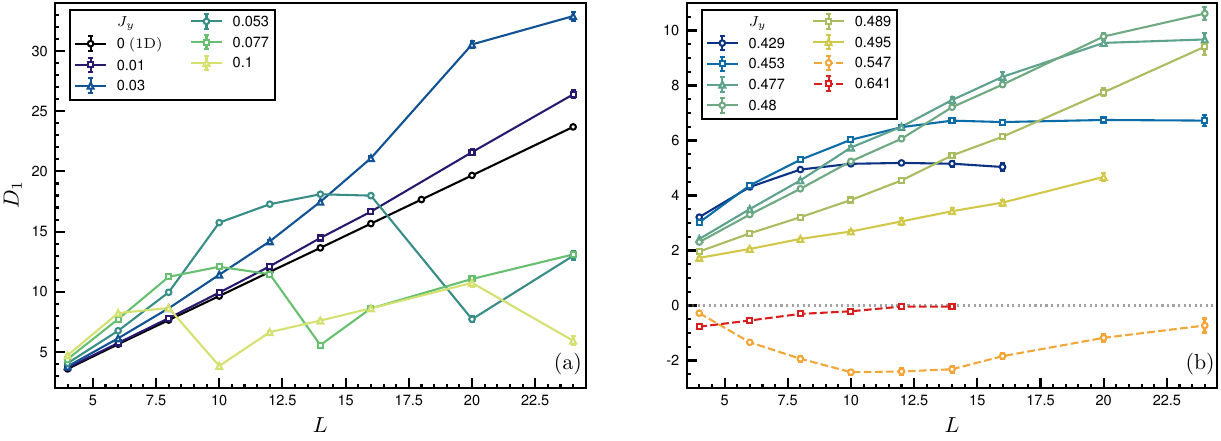}
        \caption{Volume-law correction $D_1$ (Eq.~\eqref{eq:D1}) vs $L$ (a) for low $J_y$ and (b) close to the phase transition at $J_y=0.5$.
        The results were obtained for the same system parameters used in Fig.~\ref{fig:2}, see the caption for details.}
        \label{fig:3}
    \end{figure}
    To further characterize the different system-size dependency of magic in the two phases, we extract the subleading volume-law correction $b_1$ in Eq.~\eqref{eq:vollaw} from the quantity
    \be
        \label{eq:D1}
        D_1(L)=M_1(8L^2)-4M_1(2L^2)\simeq -3b_1,
    \ee
    where $M_1(N)$ is the $1$-SRE of a system with $N$ spins.
    As shown in Fig.~\ref{fig:3}(a), our results for $D_1$ in the case of small $J_y$ are dominated by finite size effects that prohibit any further discussions.
    A comparison with the decoupled 1D limit, $J_y=0$, discussed in App.~\ref{sec:JW} demonstrates that these finite size effects start to appear for smaller sizes as $J_y$ increases deep in the gapless phase.
    On the other hand, for larger values of $J_y$ close to the phase transition, we observe the convergence of $D_1$ to its thermodynamic values $-3b_1$ in Fig.~\ref{fig:3}(b), although the convergence becomes slower close to the critical point $J_y=0.5$.
    We find that the correction $b_1$ vanishes in the gapped phase ($J_y>0.5$), while it approaches non-constant negative values in the gapless phase.

For both $m_1$ and $D_1$, we observe that the data as a function of $J_y$ for different sizes $L$ intersect at a crossing point close to $J_y=0.5$ and share a common profile, see Figs.~\ref{fig:4}(a) and \ref{fig:5}(a), respectively. 
We extract the critical exponents of $m_1$ and $D_1$ by assuming a scaling of type
\be
\label{eq:fscollapse}
\begin{split}
m_1(J_y,L) &= m_1^*+L^{y_m}f\lsb(J_y-0.5)L^{1/\nu_m}\rsb\\
D_1(J_y,L) &= D_1^*+L^{y_D}g\lsb(J_y-0.5)L^{1/\nu_D}\rsb,
\end{split}
\ee
where $f$ and $g$ are universal scaling functions and $y_m$ and $y_D$ are scaling dimensions.
By performing a Nelder-Mead minimization of the chi-square in a certain area around $J_y=0.5$ (shaded area in Figs.~\ref{fig:4}(a) and \ref{fig:5}(a)) with respect to an interpolation of the points corresponding to the largest size, we obtain $\nu_D\simeq 1$.
On the other hand, the results for the exponents of $m_1$ are found to be highly sensitive to the $J_y$ interval selected for the $\chi^2$ minimization.
In both cases, error bars are obtained by applying a bootstrap procedure which assumes each point of the largest-size curve to be distributed according to a Gaussian with standard deviation equal to the corresponding error. The rescaled curves for $m_1$ and $D_1$, for the system sizes used in the minimization, are reported in Figs.~\ref{fig:4}(b) and \ref{fig:5}(b), respectively.

We note that the result for the critical exponent $\nu\simeq 1$ is consistent with Ref.~\cite{hu2024nature}, which, in the absence of integrability-breaking perturbations, relates the anisotropy-driven topological phase transition of the KHM to the universality class of the symmetry-breaking phase transition of semi-Dirac electrons \cite{dietl2008new,montambaux2009universal,banerjee2009tight,uryszek2020fermionic}.
This can be readily understood from the anisotropic fermionic dispersion at the transition. The low-energy spectrum is quadratic along one momentum direction $a$ but linear along the orthogonal direction $b$. Consequently, the correlation lengths $\xi_a$ and $\xi_b$ are expected to exhibit distinct critical scaling, with $\xi_a\sim |J_y-0.5|^{-1/2}$ and $\xi_b\sim |J_y-0.5|^{-1}$, corresponding to the correlation-length exponents $\nu_a=1/2$ and $\nu_b=1$, respectively. Our results indicate that the critical scaling of the SRE is governed by the latter and that subleading volume-law corrections provide a more stable probe of the phase transition than the full-state magic.

Let us also stress that, in the context of a transition from QSL to a trivial state, universal collapse has also been reported~\cite{tarabunga2023many} (see also Ref.~\cite{liu2025nonequilibrium}), consistent with the correlation-length critical exponent. This may suggest that, in the continuum limit, the magic of gauge theories carries universal information.

%\tg{maybe, but I don't really know how to interpret them}
%\todo{I'm actually having some difficulties with $m_1$ because the $\chi^2$ is very large (order $10^2$) and I'm not able to reduce it. I've tried everything: removing the small sizes, changing interpolator, changing the area of $J_y$ I'm minimizing over. Nothing seems to work. Also, the results for $\nu_m$ and $y_m$ are very much dependent on the selected area of $J_y$. If it is too little or too large, $\nu_m$ can reach values between 1.5 or 2. Maybe we should just say we are not able to perform the procedure for $m_1$, but I have to say I'm embarassed. Looking at my data, the chi square is so large because the errors for m are extremely little. Despite the residues being little as well, the rescaled errors at the denominator are smaller of a factor $10$, which causes the $\chi^2$ being a sum of terms of order $100$. I guess that the situation could improve if I had smaller residues, but for that I should add a lot of points to the curve L=40 and, even if it works, I don't think I have enough time for it. I have to say that the collapsed curves look way better than the $D_1$ case for each of my attempts, but at least the results for $D_1$ are stable}
%\tb{I talked to Marcello. He said we should include it anyway.} \tg{how does he suggest to do it?} \tb{like we do now, can you draft a sentence explaining the problem for $m_1$.}
    \begin{figure}[t]
        \centering
        \includegraphics[width=\textwidth]{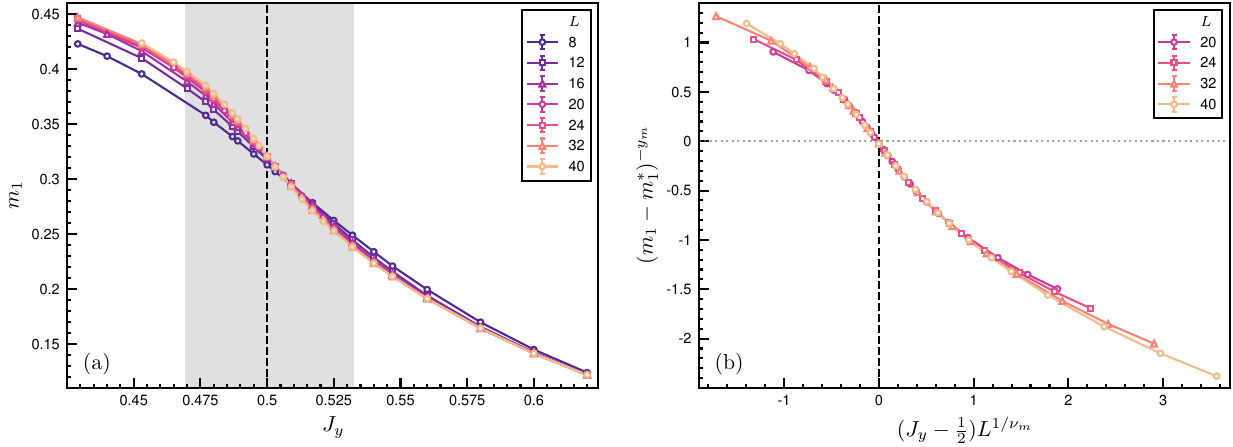}
        \caption{(a) SRE density $m_1$ around the critical point $J_y=0.5$ for different linear system sizes. The shaded area indicates the interval  used in the $\chi^2$ minimization. (b) Finite-size collapse of $m_1$ according to the rescaling in Eq.~\eqref{eq:fscollapse} for the system sizes used in the $\chi^2$ minimization, with obtained values  $m_1^*  = 0.3229 \pm 0.0001$, $y_m = -0.6701 \pm 0.0078$, $\nu_m = 1.0878 \pm 0.0122$. These values, however, strongly depend on the selected interval (shaded area in (a)). 
        The results were obtained for the same system parameters used in Fig.~\ref{fig:2}, see the caption for details.}
        \label{fig:4}
    \end{figure}
    \begin{figure}[t]
        \centering
        %\makebox[\textwidth][c]{\includegraphics[width=0.5\textwidth]{images/D1Jytr.png}\hfill
        %\includegraphics[width=0.5\textwidth]{images/D1trcollapse.png}}
        \includegraphics[width=\textwidth]{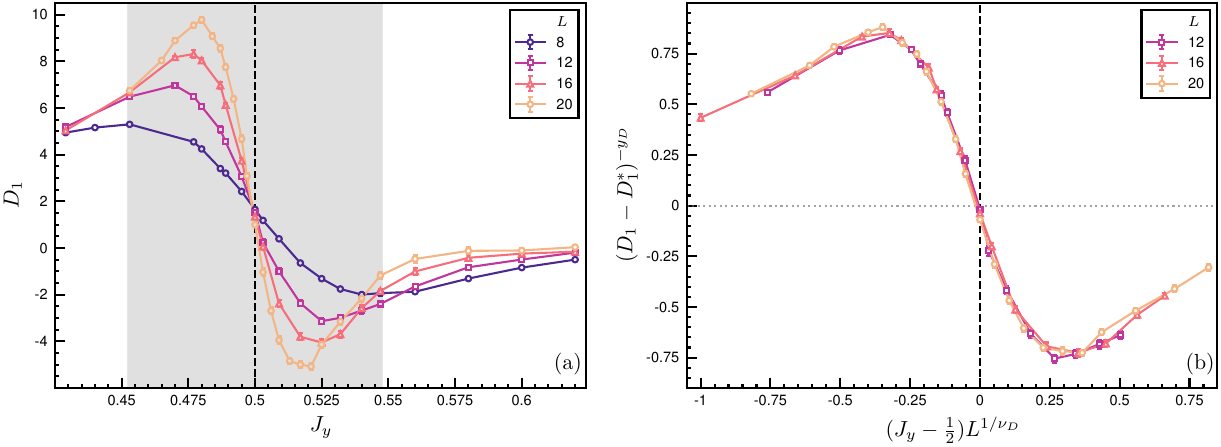}
        \caption{(a) Volume-law corrections $D_1$ vs $J_y$ close the the phase transition at $J_y=0.5$ for different linear system sizes. The shaded area indicates the interval used in the $\chi^2$ minimization. (b) Finite-size collapse of $D_1$ according to the rescaling in Eq.~\eqref{eq:fscollapse}, with obtained values $D_1^*  = 1.6396 \pm 0.1137$, $y_D = 0.7424 \pm 0.0172$, $\nu_D = 1.0484 \pm 0.0318$. The results were obtained for the same system parameters used in Fig.~\ref{fig:2}, see the caption for details.}
        \label{fig:5}
    \end{figure}

\section{Perturbation theory in the anisotropic limit}
    \label{sec:perturbation}
    \begin{figure}[t]
        \centering
        \includegraphics[width=\textwidth]{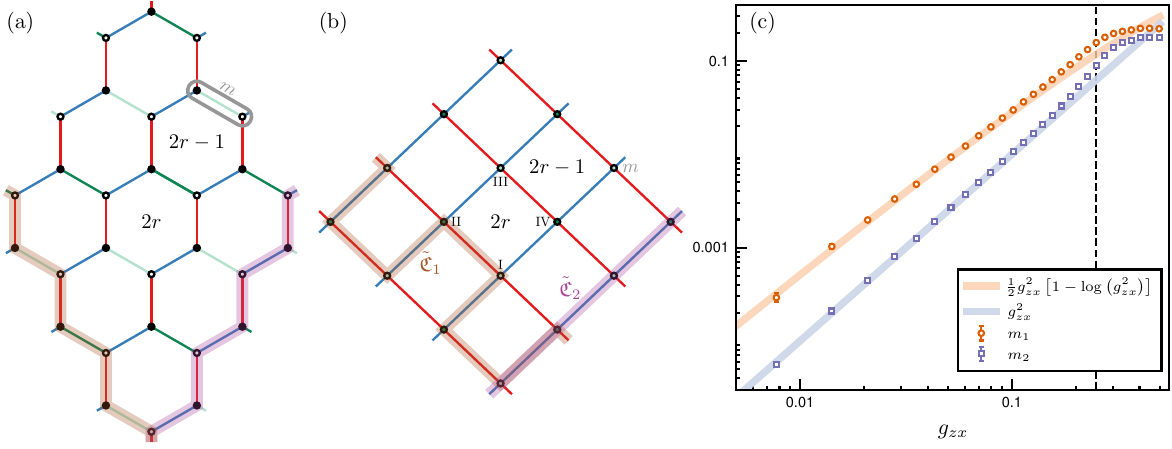}
        \caption{(a), (b) Dimers on the honeycomb lattice form an effective square lattice with square plaquettes. (c) SRE densities $m_1$ and $m_2$ as functions of the coupling parameter $g_{zx}$. Thick lines show analytical approximations from perturbation theory, while symbols indicate numerical results from the data set used in Fig.~\ref{fig:1}(c), see the caption for details. The dashed line marks the phase transition from the gapped phase ($g_{zx}<0.25$) to the gapless phase ($g_{zx}>0.25$).}
        \label{fig:6}
    \end{figure}
    To confirm our numerical results and investigate the magic structure of the gapped phase, we approximate the SRE using a perturbative expansion in the anisotropic limit $J_x=J_z\ll J_y$.
    This regime was first studied in the original work by Kitaev \cite{kitaev2006anyons}, who mapped a fourth-order expansion of the Hamiltonian~\eqref{eq:hamiltonian} to the toric code model.
    A systematic high-order expansion based on \emph{perturbative continuous unitary transformations} (PCUTs) \cite{knetter2000perturbation} was performed in Refs.~\cite{dusuel2008creation,schmidt2008emergent,vidal2008perturbative}, and showed that the ground state maps to the toric-code ground state at \emph{any order} beyond the fourth order.
    These results imply that the ground state is a stabilizer state in a rotated basis {(obtained from the PCUT)} at every order,\footnote{This is because the mapping to the toric-code ground state is a Clifford transformation \cite{kitaev2006anyons,vidal2008perturbative}.} and that the finite SRE observed in our numerical results arises from the rotation of the Pauli strings defined in the original basis.
    In particular, it turns out that we can obtain the lowest-order estimate of the SRE by rotating the Pauli strings using the unitary transformation corresponding to a second-order expansion.
    In the following, we thus compute the expectation values of the \emph{second-order-expanded} Pauli strings with respect to the \emph{high-order-expanded} ground state.\footnote{We note that, for a purely numerical comparison, the expansion of the ground state beyond second order is, strictly speaking, not necessary. However, we use this state to capture the correct spatial magic structure and facilitate generalizations in future works.}

\subsection{Stabilizer ground state}
    In the following, we briefly review and adapt the formalism introduced in Ref.~\cite{vidal2008perturbative}, see also Refs.~\cite{dusuel2008creation,schmidt2008emergent} for details.
    To this end, we rewrite the Hamiltonian~\eqref{eq:hamiltonian} as
    \be
        \label{eq:rescaled_hamiltonian}
        H=H_y+H_{zx},\qquad H_y=-\frac{1}{2}\sum_{\braket{j,k}_y}\sigma^y_j\sigma^y_k,\qquad H_{zx}=-g_{zx}\sum_{\braket{j,k}_x}\sigma^x_j\sigma^x_k-g_{zx}\sum_{\braket{j,k}_z}\sigma^z_j\sigma^z_k,
    %H_{zx}=-g_{zx}\lsb\sum_{\braket{j,k}_x}\sigma^x_j\sigma^x_k+\sum_{\braket{j,k}_z}\sigma^z_j\sigma^z_k\rsb,
    \ee
    where we set $J_y=0.5$ and $g_{zx}=J_z=J_x$ for convenience.
    Assuming $0\leq g_{zx}\ll 1$, we can then treat $H_{zx}$ as a perturbation to $H_y$.
    The unperturbed system, with $g_{zx}=0$, has an extensive ground-state degeneracy, and its ground-state manifold is spanned by ferromagnetic spin pairs on the $y$-bonds, i.e., product stabilizer states of the form $\bigotimes_m \ket{\sigma_{2m-1}^y=s_m}\ket{\sigma_{2m}^y=s_m}$, where $\mb s\in\{-1, +1\}^{N_\mr p}$.
    These spin-pair dimers form an effective square lattice, see Figs.~\ref{fig:6}(a), (b), and are conveniently described within the representation
    \begin{subequations}
    \be
    \label{eq:pertrep1}
        \sigma_{2m-1}^x=b_m+b_m^\dagger,\qquad \sigma_{2m-1}^y=\tau^z_m(-1)^{n_m},\qquad \sigma_{2m-1}^z=-i\tau^z_m(b_m-b_m^\dagger),
    \ee
    \be
        \label{eq:pertrep2}
        \sigma_{2m}^x=\tau^x_m(b_m+b_m^\dagger),\qquad \sigma_{2m}^y=\tau^z_m,\qquad \sigma_{2m}^z=-\tau^y_m(b_m+b_m^\dagger),
    \ee
    \end{subequations}
    where $b_m$ ($b_m^\dagger$) is the annihilation (creation) operator of a hardcore boson on the dimer $m$, $n_m=b_m^\dagger b_m$ is the boson density and the Pauli operators $\tau^\alpha_m$, with $\alpha\in\{x,y,z\}$, act on a spin-$1/2$ degree of freedom on the dimer $m$.
    A boson on dimer $m$ corresponds to an antiferromagnetic alignment of the corresponding spin pair, and thus describes an excitation of energy $\varepsilon=1$.
    Within this representation, the Hamiltonian~\eqref{eq:rescaled_hamiltonian} reads
    \be
        H=N_b-\frac{1}{2}N_\mr p+T_0+T_2+T_{-2},
    \ee
    where $N_b=\sum_m n_m$ is the boson number, $N_\mr p=N/2$ is the number of dimers and we defined the hopping terms
    \begin{subequations}
        \be
            T_0=-g_{zx}\sum_{m=1}^{N_\mr p}\lsb\tau^x_{q_{xm}}\lb b_m^\dagger b_{q_{xm}}+\mr{h.c.}\rb-i\tau^z_m\tau^y_{q_{zm}}\lb b_m^\dagger b_{q_{zm}}-\mr{h.c.}\rb\rsb,
        \ee
        \be
            T_2=-g_{zx}\sum_{m=1}^{N_\mr p}\lb\tau^x_{q_{xm}} b_m^\dagger b_{q_{xm}}^\dagger-i\tau^y_m\tau^z_{q_{zm}}b^\dagger_mb_{q_{zm}}^\dagger\rb,\quad T_{-2}=T_2^\dagger.
        \ee
    \end{subequations}
    Here, $q_{\alpha m}$ denotes the dimer connected to $m$ by the bond $\braket{2m-1, 2q_{\alpha m}}_\alpha$.
    The hopping terms satisfy the relation $[N_b,T_r]=rT_r$ for $r\in\{-2,0,2\}$, and thus meet the requirements for the general perturbative expansion developed in Ref.~\cite{knetter2000perturbation}.
    In short, this expansion is based on a unitary transformation $U_{\ell}$ that eliminates all terms that change the boson number up to a desired order $g_{zx}^\ell$ for an integer $\ell\geq 0$.
    The effective low-energy Hamiltonian is then given by $H_\eff=\Pi_0U_{\ell}^\dagger HU_{\ell}\Pi_0$, where $\Pi_0=\prod_m(1-n_m)$ is the projector onto the boson-free subspace.
    This Hamiltonian is naturally expressed in terms of the plaquette operators~\eqref{eq:plaquette} and Wilson loops~\eqref{eq:wilson}, which {within the representation defined by Eqs.~\eqref{eq:pertrep1} and ~\eqref{eq:pertrep2}}, transform to
    \be
        \label{eq:loops_square}
        W_p=-\tau^y_\mr{I}\tau^z_\mr{II}\tau^y_\mr{III}\tau^z_\mr{IV}(-1)^{n_\mr{I}+n_\mr{II}},\qquad \mf W_1=\prod_{m\in\tilde{\mf C}_1}\tau_{m}^x,\qquad \mf W_2=\prod_{m\in\tilde{\mf C}_2}
        \tau^z_m(-1)^{(m\;\mr{mod}\;2)n_m},
    \ee
    using the convention indicated in Fig.~\ref{fig:6}(b).
    At fourth order, $\ell=4$, the effective Hamiltonian couples to the plaquette operators and thereby lifts the extensive degeneracy up to a residual four-fold topological ground-state degeneracy \cite{kitaev2006anyons,vidal2008perturbative}.
    The latter degeneracy is partially or fully lifted by corrections that couple to the non-contractible Wilson loops $\mf W_1$ and $\mf W_2$ and appear at orders above linear system size ($\ell\geq L$) \cite{vidal2008perturbative}.
    
    For the global ground state, we can circumvent the construction of the high-order effective Hamiltonian by specifying the ground-state stabilizer structure.
    To this end, note that since the ground state $\ket{\Psi_\eff}$ of $H_\eff$ always lives in the boson-free subspace, i.e., $\Pi_0\ket{\Psi_\eff}=\ket{\Psi_\eff}$, it is stabilized by the $y$-bond operators,
    \be
        \sigma^y_{2m-1}\sigma^y_{2m}\ket{\Psi_\eff}=\ket{\Psi_\eff},\qquad m\in\{1,...,N_\mr p\}.
    \ee
    Combining these stabilizers with the stabilizers of the flux sectors, see Eq.~\eqref{eq:stabilizer_flux}, we infer that the stabilizer group $\mc S_\eff=\braket{\{\mc G_\eff\}}$ of $\ket{\Psi_\eff}$ is generated by the $2N_\mr p$ independent generators
    \be
        \label{eq:stabilizer_eff}
        \mc G_{\eff }=\{w_1W_1,\; w_2W_2,\; ...,\; w_{N_\mr p-2}W_{N_\mr p-2},\;\mf w_1\mf W_1,\;\mf w_2\mf W_2,\;\sigma_1^y\sigma_2^y,\;\sigma_3^y\sigma_4^y,\;...,\;\sigma^y_{2N_\mr p-1}\sigma^y_{2N_\mr p}\},
    \ee
    where $w_1=...=w_{N_\mr p}=1$ and $(\mf w_1,\mf w_2)\in\{+1,-1\}^2$ parametrize the vortex-free sector with lowest energy, see Sec.~\ref{sec:model}.
    To verify that these generators are indeed independent, note that within the boson-free sector, plaquette operators~\eqref{eq:loops_square} \emph{on different sublattices of the dual lattice} are subject to distinct $\mathbb Z_2$ constraints \cite{kitaev2006anyons},
    \be
        \prod_{r=1}^{N_\mr p/2} \Pi_0W_{2r-1}\Pi_0=\mathbb 1,\qquad  \prod_{r=1}^{N_\mr p/2} \Pi_0W_{2r}\Pi_0=\mathbb 1,
    \ee
    where we use the notation indicated in Fig.~\ref{fig:6}(b).
    Since $\mc S_\eff$ has $N=2N_\mr p$ independent generators and the Hilbert space dimension is $2^N$, it follows by virtue of the stabilizer formalism that $\ket{\Psi_\eff}$ is a uniquely specified stabilizer state~\cite{gottesman1997stabilizer,nielsen2010quantum}.
    Moreover, for appropriate values of $\mf w_1$ and $\mf w_2$, it is the ground state of $H_\eff$ at \emph{any} order in perturbation theory \cite{vidal2008perturbative}.
        
\subsection{Schrieffer-Wolff transformation of Pauli strings}
    A technical advantage of using PCUTs $U_\ell$ is the straightforward computation of expectation values of observables $O$ in terms of $\braket{\Psi_\eff|O_\eff|\Psi_\eff}$ where $O_\eff=U_\ell^\dagger OU_\ell$ is the rotated operator\cite{knetter2000perturbation,vidal2008perturbative}.
    In our case, it suffices to expand up to second order and employ the canonical Schrieffer-Wolff transformation $U_2=\mr e^{-\Omega}$ with generator
    \be
        \Omega=\Omega_1+\Omega_2,\qquad \Omega_1=T_{2}-T_{-2},\qquad \Omega_2=\frac{1}{4}\lsb T_2+T_{-2},T_0\rsb,
    \ee
    where $\Omega_1$ and $\Omega_2$ are operators of first and second order in $g_{zx}$, respectively \cite{macdonald1988t,chernyshev2004higher}.
    The generator is constructed such that the Hamiltonian conserves the boson number $N_b$ up to order $g_{zx}^2$.
    In the boson-free subspace, the perturbative correction then merely amounts to a constant energy shift,
    \be
        H_\eff=\Pi_0 U_{2}^\dagger HU_{2}\Pi_0=-\frac{1}{2}\lb 1-2g_{zx}^2\rb N_\mr p+\mc O\lb g_{zx}^3\rb.
    \ee
    Non-trivial corrections can, however, arise in the expansion of an arbitrary Pauli string $\bs\sigma\in\mc P$,
    \be
        \mr{e}^\Omega\bs\sigma\mr{e}^{-\Omega}=\bs \sigma+[\Omega_1,\bs \sigma]+\frac{1}{2}[\Omega_1,[\Omega_1,\bs \sigma]]+[\Omega_2,\bs \sigma]+\mc O\lb g_{zx}^3\rb.
    \ee
    To compute the $n$-SRE~\eqref{eq:sre}, we need the $2n$th powers of expectation values of Pauli strings.
    Using the shorthand notation $\braket{...}=\braket{\Psi_\eff|...|\Psi_\eff}$, we obtain
    \be
        \label{eq:trafo_pauli}
        \begin{split}
            %\sum_{\bs\sigma\in\mc P}\braket{\mr e^{\Omega}\bs \sigma\mr e^{-\Omega}}^{2n}&=\sum_{\bs\sigma\in\mc P}\lsb\braket{\bs \sigma}+\binom{2n}{1}\braket{\bs\sigma}^{2n-1}\braket{[\Omega_1,\bs\sigma]}+\binom{2n}{2}\braket{\bs\sigma}^{2n-2}\braket{[\Omega_1,\bs\sigma]}^2\right.\\
            %&\left.+\binom{2n}{1}\braket{\bs\sigma}^{2n-1}\braket{\frac{1}{2}[\Omega_1,[\Omega_1,\bs\sigma]]+[\Omega_2,\bs\sigma]}+\mc O\lb g^{3}_{zx}\rb\rsb
            \braket{\mr e^{\Omega}\bs \sigma\mr e^{-\Omega}}^{2n}&=\braket{\bs \sigma}^{2n}+\binom{2n}{1}\braket{\bs\sigma}^{2n-1}\braket{[\Omega_1,\bs\sigma]}+\binom{2n}{2}\braket{\bs\sigma}^{2n-2}\braket{[\Omega_1,\bs\sigma]}^2\\
            &+\binom{2n}{1}\braket{\bs\sigma}^{2n-1}\braket{\frac{1}{2}[\Omega_1,[\Omega_1,\bs\sigma]]+[\Omega_2,\bs\sigma]}+\mc O\lb g^{3}_{zx}\rb.
        \end{split}
    \ee
    \begin{figure}
        \centering
        \includegraphics[width=\columnwidth]{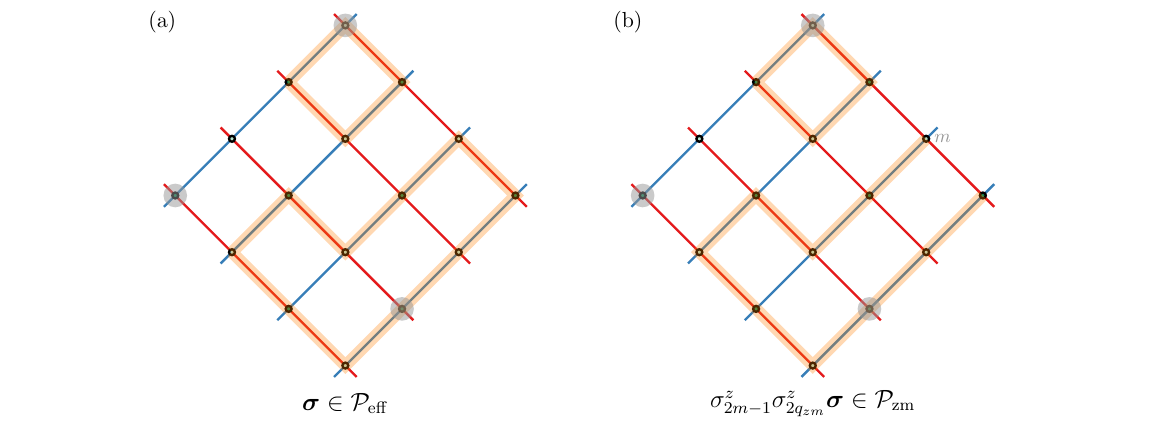}
        \caption{{Graphical representation of Pauli strings on the square lattice. Thick orange lines correspond to perturbative bond operators $\sigma_{2m-1}^\alpha\sigma^\alpha_{2q_{\alpha m}}$, with $\alpha\in\{x,z\}$, and thick gray circles mark on-dimer operators $\sigma^y_{2m-1}\sigma^y_{2m}$. (a) Pauli strings in $\mc P_\eff$ involve only closed loops and correspond (up to a sign) to a ground-state stabilizer. (b) Pauli string in $\mc P_{zm}$ with an open world line terminating at $m$ and $q_{zm}$.}}
        \label{fig:7}
    \end{figure}
    \paragraph*{R\'enyi coefficient $n=2$.}
    Considering the R\'enyi coefficient $n=2$, we notice that every term in the expansion involves a factor $\braket{\bs\sigma}$, which, given that $\ket{\Psi_\eff}$ is a stabilizer state, is nonzero only if $-\bs\sigma$ or $\bs\sigma$ is a stabilizer.
    We can therefore restrict the summation over the set of all Pauli strings, $\mc P$, to the summation over the set
    \be
        \mc P_\eff=\{\bs\sigma\in\mc P\mid \bs\sigma\in\mc S_\eff\text{ or }-\bs\sigma\in\mc S_\eff\},
    \ee
    {which, from a physical perspective, comprises the Pauli strings that generate boson-pair creation-annihilation processes along closed loops or probe the local bosonic parities, see Fig.~\ref{fig:7}(a).}
    Since all stabilizers in Eq.~\eqref{eq:stabilizer_eff} conserve the boson number $N_b$, we can then eliminate all terms that change $N_b$, yielding\footnote{More specifically, we obtain $\braket{\bs\sigma}^3\braket{[\Omega_1,[\Omega_1,\bs\sigma]]}=\frac{1}{4}\braket{\bs\sigma}^3\braket{ 2T_{-2}\bs\sigma T_2-\bs\sigma T_{-2}T_2-T_{-2}T_2\bs\sigma}$ and $\braket{\bs\sigma}\braket{\lsb\Omega_1,\bs\sigma\rsb}=\braket{\bs\sigma}\braket{\lsb\Omega_2,\bs\sigma\rsb}=0$. }
    \be
        \label{eq:sum_n2}
        \sum_{\bs\sigma\in\mc P}\braket{\mr e^{\Omega}\bs \sigma\mr e^{-\Omega}}^{4}\simeq\sum_{\bs\sigma\in\mc P_\eff}\lb\braket{\bs\sigma}^4+\frac{1}{2}\braket{\bs\sigma}^3\braket{ 2T_{-2}\bs\sigma T_2-\bs\sigma T_{-2}T_2-T_{-2}T_2\bs\sigma}\rb.
    \ee
    Upon summing over $\mc P_\eff$, the first term in the correction gives a vanishing contribution
    \be
        \sum_{\bs\sigma\in \mc P_\eff}\braket{\bs\sigma}^3\braket{T_{-2}\bs\sigma T_2}=g_{zx}^2\sum_{\bs\sigma\in \mc P_\eff}\sum_{\alpha\in\{x,z\}}\sum_{m=1}^{N_\mr p}\braket{\bs\sigma}^3\braket{\sigma^\alpha_m\sigma^\alpha _{q_{\alpha m}}\bs\sigma\sigma^\alpha_m\sigma^\alpha_{q_{\alpha m}}}=0.
    \ee
     To understand the second equality, notice that a given bond operator $\sigma^\alpha_m\sigma^\alpha_{q_{\alpha m}}$ anticommutes with $2$ generators in Eq.~\eqref{eq:stabilizer_eff}, while it commutes with all other $N-2$ generators, and therefore
     \be
        \sum_{\bs\sigma\in\mc P_\eff}\braket{\bs\sigma}^3\braket{\sigma^\alpha_m\sigma^\alpha_{q_{\alpha m}}\bs \sigma\sigma^\alpha_m\sigma^\alpha_{q_{\alpha m}}}=2^{N-1}-2^{N-1}=0.
    \ee
    The remaining two corrections in Eq.~\eqref{eq:sum_n2} merely renormalize the leading contribution and yield $\braket{\bs\sigma}^3\braket{\bs \sigma T_{-2}T_2}=\braket{\bs\sigma}^3\braket{T_{-2}T_2\bs \sigma  }=Ng_{zx}^2\braket{\bs\sigma}^4$.
    Together with Eq.~\eqref{eq:sre} and the identity $|\mc P_\eff|=2^N$, we then obtain the $2$-SRE
    \be
        \label{eq:perturb_sre2}
        M_2\simeq-\log\lb 1-Ng_{zx}^2\rb\simeq Ng_{zx}^2.
    \ee
        \paragraph*{R\'enyi coefficient $n=1$.} While the generalization to higher R{\'e}nyi indices $n\geq 2$ is straightforward, the calculation is more subtle in the case of the $1$-SRE,
    \be
        \label{eq:sre1}
        M_1=-\frac{1}{2^N}\sum_{\bs\sigma\in\mc P}\braket{\mr e^{\Omega}\bs \sigma\mr e^{-\Omega}}^2\log{\lb\braket{\mr e^{\Omega}\bs \sigma\mr e^{-\Omega}}^2\rb}.
    \ee
    This is because the expansion~\eqref{eq:trafo_pauli} for $n=1$ involves the term $\braket{\lsb\Omega_1,\bs\sigma\rsb}^2$, which is not restricted to $\bs\sigma\in\mc P_\eff$.
    Instead, we find
    \be
        \braket{[\Omega_1,\bs\sigma]}=\frac{g_{zx}}{2}\sum_m\lb\braket{\{\sigma^x_{2m-1}\sigma^x_{2q_{xm}},\bs\sigma\}}+\braket{\{\sigma^z_{2m-1}\sigma^z_{2q_{zm}},\bs\sigma\}}\rb,
    \ee
    where $\{...,...\}$ denotes the anticommutator.
    Since $\ket{\Psi_\eff}$ is a stabilizer state, we infer that a given term $\braket{\{\sigma^\alpha_{2m-1}\sigma^\alpha_{2q_{\alpha m}},\bs\sigma\}}$ is only finite for Pauli strings $\bs\sigma\in\mc P_{\alpha m}$, where
    \be
        \mc P_{\alpha m}=\{\sigma^\alpha_{2m-1}\sigma^\alpha_{2q_{\alpha m}}{\bs\sigma}\mid {\bs\sigma}\in\mc P_\eff\text{ and }[\sigma^\alpha_{2m-1}\sigma^\alpha_{2q_{\alpha m}},{\bs\sigma}]=0\}\subset\mc P.
    \ee
    {Physically, $\mc P_{\alpha m}$ comprises Pauli strings that generate boson-pair creation-annihilation processes including an open world line corresponding to the creation of bosons on the dimers $m$ and $q_{m\alpha}$, see Fig.~\ref{fig:7}(b)}.}
    We note that the sets $\mc P_{\alpha m}$ are mutually disjoint and disjoint from $\mc P_\eff$, and thus allow us to partition the summation in Eq.~\eqref{eq:sre1}.
    Using the same arguments as above, we find that the sum over $\mc P_\eff$ yields a quadratic contribution,
    \be
        \sum_{\bs\sigma\in\mc P_\eff}\braket{\mr{e}^\Omega\bs\sigma\mr{e}^{-\Omega}}^2\log\lb\braket{\mr{e}^\Omega\bs\sigma\mr{e}^{-\Omega}}^2\rb=-2^N\lb 1-\frac{N}{2}g_{zx}^2\rb\log{\lb 1-\frac{N}{2}g_{zx}^2\rb}\simeq -2^N\frac{N}{2}g_{zx}^2,
    \ee
    while the term $\braket{\lsb\Omega_1,\bs\sigma\rsb}^2$ generates the contribution (using $|\mc P_{\alpha m}|=2^{N-1}$)
    \be
        \sum_{m=1}^{N_\mr p}\sum_{\alpha\in\{x,z\}}\sum_{\bs\sigma\in\mc P_{\alpha m}}\braket{\mr{e}^\Omega\bs\sigma\mr{e}^{-\Omega}}^2\log\lb\braket{\mr{e}^\Omega\bs\sigma\mr{e}^{-\Omega}}^2\rb=2^{N}Ng_{zx}^2\log{\lb g_{zx}^2\rb}.
    \ee
    Collecting all terms yields the $1$-SRE
    \be
        \label{eq:perturb_sre1}
        M_1%=-\frac{1}{2^N}\sum_{\bs\sigma\in\mc P}\braket{\bs\sigma}^2\log{\lb\braket{\bs\sigma}^2\rb}
        \simeq\frac{N}{2}g_{zx}^2\lsb 1-\log{(g_{zx}^2)}\rsb.
    \ee
    We compare the analytical estimates of the SRE densities $m_1=M_1/N$ and $m_2=M_2/N$ with our numerical results in Fig.~\ref{fig:6}(c), and find good agreement up to $g_{zx}\sim 0.1$, followed by the breakdown of perturbation theory near the critical point $g_{zx}=0.25$.
    The results in Fig.~\ref{fig:1} are obtained using the rescaling $g_{zx}=(1-J_y)/(4J_y)$.
    Note that at no point in the above derivation did we insert the eigenvalues $w_p$, ${\mf w}_1$ and ${\mf w}_2$ of the plaquette and Wilson loop operators.
    We thus expect the approximations in Eqs.~\eqref{eq:perturb_sre2} and \eqref{eq:perturb_sre1} to remain valid in the presence of flux excitations, and, more generally, in any flux sector within the gapped phase.

\section{Conclusion and perspective}
    \label{sec:conclusion}

In this work, we investigated  magic in the Kitaev honeycomb model, providing, to the best of our knowledge, the first large-scale characterization of the stabilizer R\'enyi entropy across quantum spin liquids of different nature.

A key conceptual step was to show that the emergent Majorana fermionization underlying Kitaev's exact solution can be used not only to diagonalize the Hamiltonian, but also to access the SRE. 
In particular, by deriving a precise correspondence between physical Pauli strings and strings of itinerant Majorana operators within a fixed gauge sector, we established the equivalence between the SRE and an analogous Majorana-resolved quantity in the enlarged Hilbert space. 
This result is nontrivial because the relevant fermionization is genuinely different from the standard Jordan--Wigner mapping: while the latter does not render the physical Kitaev ground state Gaussian, the Kitaev Majorana representation exposes the free-fermion structure that makes an efficient computation possible. 
%We note that, while a solution of the Kitaev honeycomb model in terms of a Jordan–Wigner transformation is available, the procedure involves the decoupling of a quartic term, which again introduces a gauge redundancy \cite{feng2007topological}. It is therefore not obvious whether or how this or other solution schemes \cite{vidal2008perturbative,kells2009description} can enable an efficient sampling routine for the spin SRE.
%\tg{we should also mention that all these studies are done for OBC (to my knowledge)}\tb{(I added this at first but in Ref. [43] they also talk about periodic bc somewhere so i wanted to be careful (wihtout reading the whole paper again).}

On the computational side, we introduced an optimized high-performance implementation of the Gaussian-fermionic sampling algorithm, reducing the computational scaling from $O(N^4)$ to $O(N^3)$. 
This improvement, which can be readily applied to other gaussian wave functions, allowed us to access lattices with up to $N\simeq 4600$ spins and to resolve the behavior of the SRE across the full phase diagram. 
We found that the magic density is strongly enhanced in the gapless spin-liquid regime, with a maximum in the isotropic point, while it decreases upon entering the gapped phases and vanishes in the strongly anisotropic limit of decoupled dimers. 
In this limit, we supported the numerical results through a perturbative expansion, showing that the residual magic originates from the rotation of physical Pauli strings away from the stabilizer basis of the effective toric-code description. 
These results clarify how a phase that is topologically ordered and long-range entangled can nevertheless display very different degrees of nonstabilizer complexity depending on the microscopic spin basis and on the proximity to criticality. It is important to stress that the origin of magic in the gapped spin liquid phase can be directly linked to the properties of the perturbative unitary transformation relating it to a stabilizer fixed point, a fact that makes the use of such transformations in other studies of magic promising.

Beyond the extensive SRE density, we analyzed the subleading correction to the volume law. 
We found that this correction vanishes in the gapped phase, whereas it remains finite and non-constant in the gapless spin liquid. 
This behavior suggests that the gapless phase hosts a genuinely non-local component of magic, in the sense of a contribution that is not reducible to local nonstabilizer resources. 
Moreover, the same quantity displays scaling behavior at the transition between the gapless and gapped spin liquids, indicating that magic provides a sensitive probe of the continuous phase transition of the KHM. 
Taken together, these results establish the SRE as a diagnostic of many-body complexity in quantum spin liquids, complementary to entanglement-based probes and directly sensitive to the stabilizer structure of the wavefunction.

Our work opens several directions for future investigations.
First, a more detailed analysis of the ground-state Pauli spectrum in the vicinity of the transition between the gapless and gapped QSLs could provide further insight into how the anisotropic critical behavior is encoded in many-body Pauli correlations and reflected in the observed critical exponents, thereby complementing existing studies based on field theory \cite{uryszek2020fermionic,hu2024nature} and entanglement entropy \cite{yao2010entanglement,meichanetzidis2016anatomy}.

Second, since the equivalence between the SRE and its Majorana-resolved counterpart also holds for subsystems of the KHM, a result whose proof will be presented elsewhere, it would be natural to extend the present analysis of the full-state SRE to subsystem and multipartite magic measures.
This would allow one to use magic analogs of topological entanglement diagnostics and to test whether nonstabilizer resources in quantum spin liquids also organize in loop-like multipartite structures, as recently suggested for entanglement in gauge-theoretic phases \cite{lyu2025multiparty}. 
Such an extension is particularly promising for distinguishing Abelian, non-Abelian, and gapless spin liquids, and may also provide insights on the physics of other topological phases~\cite{nehra2025topological}.

Third, the present framework provides a natural starting point for investigating non-local magic in free-fermion and spin-liquid systems. 
Recent progress has shown that non-local magic can be efficiently characterized in fermionic Gaussian states~\cite{collura2026nonlocal,iannotti2026nonlocal}, suggesting that the matter sector of the Kitaev model may offer an analytically controlled setting in which to relate covariance-matrix-based diagnostics to physical spin nonstabilizerness. 
However, in the Kitaev spin liquid this connection is subtle: one must determine how the optimization over local fermionic Gaussian transformations is represented in the physical spin Hilbert space, and how gauge constraints modify the separation between local and non-local resources. 
Resolving this issue could lead to a decomposition of magic into matter and gauge contributions, in close analogy with known decompositions of the entanglement entropy in the Kitaev model \cite{yao2010entanglement,dora2018gauge}.
Finally, since our optimized algorithm applies to arbitrary fermionic Gaussian states, the computational advances developed here are not restricted to the honeycomb model. They can be directly exploited to study magic in broader classes of free-fermion systems, including disordered Kitaev models, three-dimensional Kitaev and Weyl spin liquids \cite{mandal2009exactly,eschmann2020thermodynamic,hermanns2015weyl}, amorphous lattices \cite{cassella2023exact}, spin chians~\cite{oliviero2022magic,trino2025stabilizer,khasseh2026universal,hoshino2026stabilizer}, qudits \cite{wu2009gamma}, open quantum systems \cite{shibata2019dissipative,yang2021exceptional,fukui2024magnetic}, dynamical protocols \cite{schmitt2015dynamical,kells2014topological}, and experimentally relevant quantum-simulation settings \cite{semeghini2021probing,satzinger2021realizing,kalinowski2023non,will2025probing,evered2025probing}, as well as to contribute to a broader understanding of magic in the context of quantum field theory\cite{chemissany2025infinite,benedetti2026universality,moosa2026typeiiivonneumann}.

\section*{Acknowledgements}
We thank Julien Vidal, Yukitoshi Motome, Nazzareno Africani and Riccardo Dal Molin for helpful discussions.

% TODO: include funding information
\paragraph{Funding information}
{T.~B. acknowledges funding by ERC Consolidator grant WaveNets (Grant agreement ID: 101087692).}
M.~D. was partly supported by the EU-Flagship programme Pasquans2, by the PNRR MUR project PE0000023-NQSTI, by INFN Iniziativa Specifica Quantum, and by the ERC Consolidator grant WaveNets (Grant agreement ID: 101087692). M.~C. was partially supported by the project ``QUASAR-FVG - Calcolo e Simulazione Quantistica: Sviluppo, Applicazioni e Ricerca in Friuli Venezia Giuli''.  
%\clearpage
\begin{appendix}
\numberwithin{equation}{section}

\section{Global ground state of the Kitaev honeycomb model}
\label{sec:groundstate}
\begin{figure}[t]
    \centering
    \includegraphics[width=\textwidth]{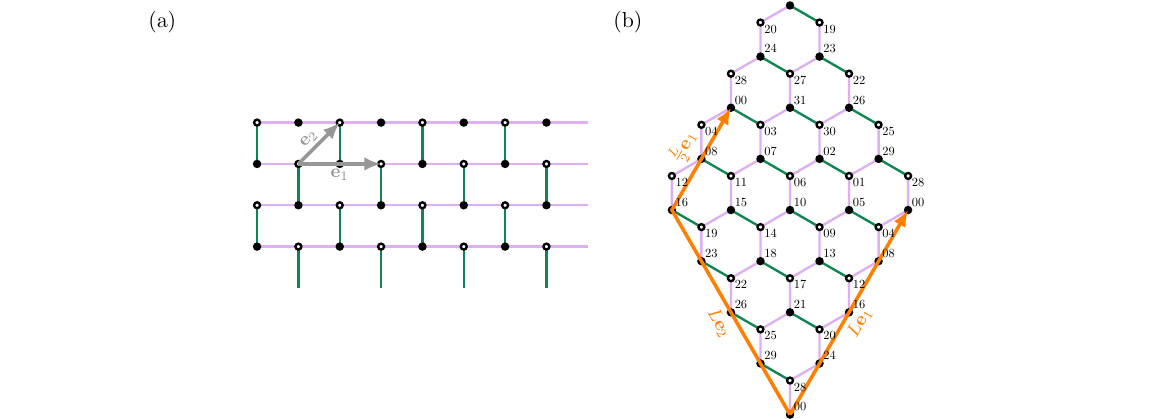}
    \caption{(a) Brick-wall lattice compatible with the geometrical requirements of Lieb's theorem for $J_x=J_z$. (b) Corresponding system on a hexagonal lattice with twisted boundary conditions defined by the supercell lattice vectors $\mb r_1 = L\mb e_1$ and $\mb r_2 = L\mb e_2+\frac{L}{2}\mb e_1$.}
    \label{fig:8}
\end{figure}

In principle, the global KHM ground state can be found by minimizing the energy over all the possible flux sectors. 
However, the computational cost of this operation grows exponentially with the system size and is thereby restricted to small systems.
This restriction can be circumvented by Lieb's theorem \cite{lieb1994flux,kitaev2006anyons}, which ensures that for a fermionic hopping model defined on a square lattice with specific periodic boundary conditions, the ground state is obtained when $W_p=-1$ for all square plaquettes $p$.
On the honeycomb lattice, this result translates into the vortex-free configuration $W_p=1$ for all hexagonal plaquettes $p$. The type of lattice for which the theorem holds has the following features. First, the sites at the right end correspond to the sites on the left, and the sites on the top correspond to the site on the {bottom}, see Fig.~\ref{fig:8}(a).
Second, the magnitudes of hopping strength along horizontal bonds are allowed to vary along each row, as long as every row is identical. 
Finally, the hopping along vertical links can vary, while moving in the horizontal direction, but must have a periodicity of 2 along the columns.
Applying these requirements to the KHM produces the brick-wall lattice in Fig.~\ref{fig:8}(b) up to spatial rotations.

Notably, the coupling strengths $J_\alpha$ and $J_\beta$ along two directions $\alpha$ and $\beta$ in Hamiltonian Eq.~\eqref{eq:hamiltonian} must have the same magnitude to satisfy the second requirement. 
The representation of the vectors $\mb e_1$, $\mb e_2$ on the brick-wall lattice, see Fig.~\ref{fig:8}(a), determines which couple of link directions gets selected in the above sense. 
The supercell lattice vectors generating the corresponding honeycomb lattice are finally defined as a function of $\mb e_1$ and $\mb e_2$ such that the periodic boundary conditions required by Lieb's theorem are satisfied. In Fig.~\ref{fig:8}(a), we show a choice for $\mb e_1$, $\mb e_2$ that is consistent with equal couplings being $J_x=J_z$ along $x$ and $z$ bonds, resulting in the green line in Fig.~\ref{fig:1}(b) studied by our numerical simulations. 
In Fig.~\ref{fig:8}(b), we show the corresponding honeycomb lattice generated by supercell lattice vectors
\begin{equation}
\label{eq:latvec}
    \mb r_1 = L_1\mb e_1,\qquad \mb r_2 = L_2\mb e_2+M\mb e_1,
\end{equation}
where $L_1=L_2=L$ and $M=L/2$.
Even if Lieb's theorem seems to only  hold for various combinations of periodic boundary conditions and parameter values, its validity in computations is typically accepted for sufficiently large system sizes. However, in general this is correct only in the gapless phase, as Lieb's theorem assumes that the ground state is a vacuum state even though, in the case of the KHM, the vacuum might not have the fermionic parity required for physical states, see Eq.~\eqref{eq:pphys}.
While this is easily solved in the gapless phase by adding a zero-energy fermion, this problem can persist in the gapped phase even for large system sizes \cite{zschocke2015physical}.

In our work we avoid finite-size effects by computing SREs for points of the phase diagram where Lieb's theorem rigorously applies and for a lattice geometry with appropriate boundary conditions.
The global ground state is found by minimizing the vacuum ground state over the four vortex-free flux sectors, and then checking if the required parity according to Eq.~\eqref{eq:pphys} is even. 
The geometric parameter $\theta$ in Eq.~\eqref{eq:pphys} is defined as \cite{pedrocchi2011physical,zschocke2015physical}
\begin{equation}
    \theta = L_1+L_2+M(L_1-M),
\end{equation}
which gives $\theta=2L+L/2(1-L/2)$ in our case, see Fig.~\ref{fig:8}(b).
\section{Limit of decoupled spin chains}
\label{sec:JW}
If the coupling strengths associated to one bond type $\alpha\in\{x,y,z\}$ all vanish, i.e., $J_\alpha=0$, the KHM reduces to a set of decoupled periodic 1D spin chains.
This limit can be diagonalized using a standard Jordan-Wigner transformation (without gauge redundancy) for a single chain. 
For the computation of the SRE, this simplification allows for both larger linear system sizes $L$ and larger sample sizes of order $N_\mr s\sim 10^{{7}}$, thereby providing a valuable benchmark of the numerical results in the main text as well as high-quality results for SRE along the solid black lines in the phase diagram of Fig.~\ref{fig:1}(b).

\subsection{Jordan-Wigner transformation}
To employ standard notation, let us assume $J_z=0$ and focus on a chain of $L$ spins subject to the Hamiltonian
\begin{equation}
\label{eq:1dham}
    H = \sum_{j=1}^{L/2}(J_y\sigma^y_{2j-1}\sigma^y_{2j} + J_x\sigma^x_{2j}\sigma^x_{2j+1}),
\end{equation}
where we impose the periodic boundary conditions $ \sigma^{{x}}_{L+1} = \sigma^{{x}}_1$.
A standard Jordan-Wigner transformation of spins into complex Dirac fermions $a_j$ yields
\begin{equation}
\label{eq:JW}
\begin{split}
    &\sigma^x_{j}\sigma^x_{j+1} = (a^{\dagger}_j-a_j)(a^{\dagger}_{j+1}+a_{j+1})\\
    &\sigma^y_{j}\sigma^y_{j+1} = -(a^{\dagger}_j+a_j)(a^{\dagger}_{j+1}-a_{j+1})\\
    &\sigma^x_{L}\sigma^x_{1} = -e^{i\pi L}(a^{\dagger}_L-a_L)(a^{\dagger}_{1}+a_{1}).
    \end{split}
\end{equation}
Treating fermions on odd and even sites as two distinct species, $A_j = a_{2j-1}$, $B_j = a_{2j}$, we arrive at
\begin{equation}
    H = \sum_{j=1}^{L/2}\lsb J_y(A^\dagger_j+A_j)(-B^{\dagger}_j+B_j)+J_x(B^\dagger_j-B_j)(A^{\dagger}_{j+1}+A_{j+1})-J_x \mr e^{i\pi L}(B^{\dagger}_L-B_L)(A^{\dagger}_1+A_1)\rsb.
\end{equation}
A Fourier transformation should take into account that the boundary term in \eqref{eq:JW} accumulates a sign $+1$ if the parity is odd and $-1$ if the parity is even, which implies PBCs in the first case and APBCs in the second. This, in turn, defines the associated operators in momentum space,
\begin{equation}
    A_j = \sqrt{\frac{2}{L}}\sum_k \mr e^{i k j}\tilde{A}_k,\quad B_j = \sqrt{\frac{2}{L}}\sum_k \mr e^{i k j}\tilde{B}_k,
\end{equation}
where we use the momentum grid
\begin{equation}
k=
\begin{cases}
\dfrac{4n\pi}{L},
& n=-\dfrac{L}{4}+1,\ldots,\dfrac{L}{4},
\qquad \qquad \ \ \text{odd parity},
\\[1ex]
\pm\dfrac{2(2n-1)\pi}{L},
& n=1-\dfrac{1}{2}\bigl(\frac{L}{2}\bmod 2\bigr),\ldots,\dfrac{L}{4},
\quad \text{even parity}.
\end{cases}
\end{equation}
By defining $\Psi_k = (\tilde{A}_k, \tilde{B}_k, \tilde{A}^{\dagger}_{-k},\tilde{B}^{\dagger}_{-k})^T$ we obtain the expression for the Hamiltonian \eqref{eq:1dham} in Fourier representation, $H = \sum_k\Psi_k^{\dagger}H_k\Psi_k$, where
\begin{equation}
    H_k = \frac{1}{2}\begin{pmatrix}
0 & \xi_k & 0 & -\xi_k\\
\xi_k^* & 0 & \xi_k^* & 0 \\
0 & \xi_k & 0 & -\xi_k\\
-\xi_k^* & 0 & -\xi_k^* & 0 
\end{pmatrix},
\end{equation}
and $\xi_k = J_y+J_x \mr e^{-i k}$.
The Hamiltonian is diagonalized by the Bogoliubov transformation (preserving anticommutation relations) 
$W_k=\bmat 
        \mb v_1(k) & \mb v_2(k) & \mb v_3(k) & \mb v_4(k)
    \emat$
with column vectors
\be
    \mb v_1(k) = 
    \frac{1}{2}
    \begin{pmatrix}
    \mr e^{i \phi_k} \\ -1 \\ \mr e^{i \phi_k} \\ 1
    \end{pmatrix}, \quad
    \mb v_2(k) = \frac{1}{2}
    \begin{pmatrix}
    -1 \\ 1 \\ 1 \\ 1
    \end{pmatrix}, \quad
    \mb v_3(k) = \frac{1}{2}\begin{pmatrix}
    1 \\ 1 \\ -1 \\ 1
    \end{pmatrix}, \quad
    \mb v_4(k) = \frac{1}{2}
    \begin{pmatrix}
    \mr e^{i \phi_k} \\ 1 \\ \mr e^{i \phi_k} \\ -1
    \end{pmatrix},
\ee
where we defined
\begin{equation}
\label{eq:phase}
\mr e^{i \phi_k} = \frac{\xi_k}{|\xi_k|} = \frac{J_y+J_x \mr e^{-i k}}{J_x^2+J_y^2+2 J_x J_y \cos{k}}.
\end{equation}
We obtain $W_k^\dagger H_kW_k=\mr{diag}\lb-\varepsilon(k), 0, 0, \varepsilon(k)\rb$, where dispersion relation relation is given by
\be
    \varepsilon(k)=\sqrt{J^2_x+J^2_y+2 J_x J_y \cos{k}}.
\ee
We note that the band gap closes at the Brillouin zone boundary, $k=\pi$, for $J_x=J_y$, indicating the tricritcal point between the two gapped phases and the gapless one in Fig.~\ref{fig:1}(b).\footnote{It is understood that the denominator of Eq.~\eqref{eq:phase} vanishes for $J_x = J_y$ and $k=\pi$. However, the point $k=\pi$ is associated to odd parity, while the ground state in our numerics always corresponds to the vacuum state with even parity.}

\subsection{Numerical results}
\begin{figure}[t]
%\vspace*{-1.2cm}
%\makebox[\textwidth][c]{\includegraphics[width=0.25\textwidth]{images/m1conv.png}
%\includegraphics[width=0.25\textwidth]{images/m1jx (1).png}
%\includegraphics[width=0.25\textwidth]{images/d1conv1d.png}
%\includegraphics[width=0.25\textwidth]{images/d1jx (1).png}}
%\makebox[\textwidth][c]{\includegraphics[width=0.25\textwidth]{images/m2conv.png}
%\includegraphics[width=0.25\textwidth]{images/m2jx (1).png}
%\includegraphics[width=0.25\textwidth]{images/d2conv1d.png}
%\includegraphics[width=0.25\textwidth]{images/d2jx (1).png}}
    \centering
    \includegraphics[width=\textwidth]{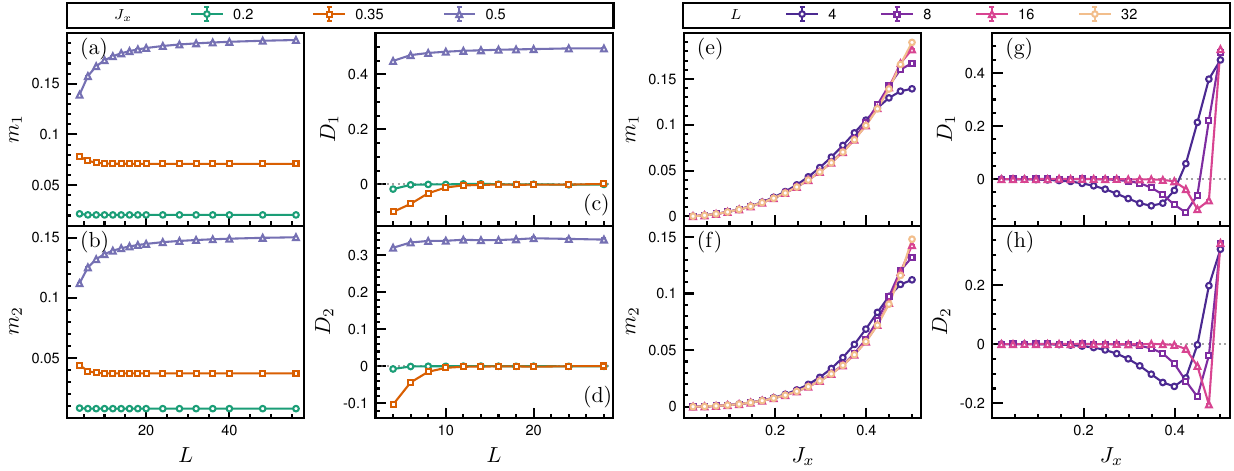}
    \caption{Numerical results for the 1D limit, $J_z=0$, corresponding to decoupled periodic spin chains. The SRE densities (a), (e) $m_1$ and (b), (f) $m_2$ were obtained for spin chains with length $2L$ along the line $J_x=1-J_y$, see Fig.~\ref{fig:1}(b), and for the samples size $N_\mr s=10^7$. The corresponding volume-law corrections (c), (g) $D_1$ and (d), (h) $D_2$ were computed from Eq.~\eqref{eq:D11d}.}
    \label{fig:9}
\end{figure}

Fig.~\ref{fig:9} includes the results for the $1$- and $2$-SRE density of a chain with parameters $J_z=0$, $J_x=1-J_y$ and
$L$ unit cells, i.e., $2L$ spins. 
Since it is possible to collect a large number of samples compared to the linear system size $L$ in this limit, we are able to converge $m_2$ in $L$, see Fig.\ref{fig:9} (a), while this remains challenging in the coupled 2D system, see App. \ref{sec:m2}. 
We observe that both $m_1$ and $m_2$
increase as $J_x$ approaches the critical value $J_x=0.5$, see Fig.~\ref{fig:9}(b), and that this effect is more pronounced for larger sizes. 
Also, the curves along $J_x$ are steeper for $m_2$, as already observed in the 2D system, see Fig.~\ref{fig:1} (c).
The assumed volume law for the SRE in the 1D system
\be
    M_{n}(L)\sim 2a_{n}L + b_{n}
\ee
allows us to extract the subleading correction using
\be
\label{eq:D11d}
D_{n}(L)=M_{n}(2L)-2M_{n}(L)\sim -b_n.
\ee
We report that $D_1$ and $D_2$ both converge with $L$, as shown in Fig.~\ref{fig:9}(c) for some values of $J_x$. 
While the convergence to zero is fast for $J_x\neq 0.5$, we observe a slow convergence to a finite value at the critical point $J_x=0.5$. This is confirmed by Fig.~\ref{fig:9}(d), in which we observe an evident transition of both $D_1$ and $D_2$ from zero to a finite value,  which becomes steeper for growing system size. 
This observation validates our two-dimensional results in the main text, see Sec.~\ref{sec:num}, where we also find a vanishing volume-law correction in the gapped phase.

\section{Numerical results for the 2-stabilizer R{\'e}nyi entropy}
\label{sec:m2}
\begin{figure}[t]
    \centering
    \includegraphics[width=\textwidth]{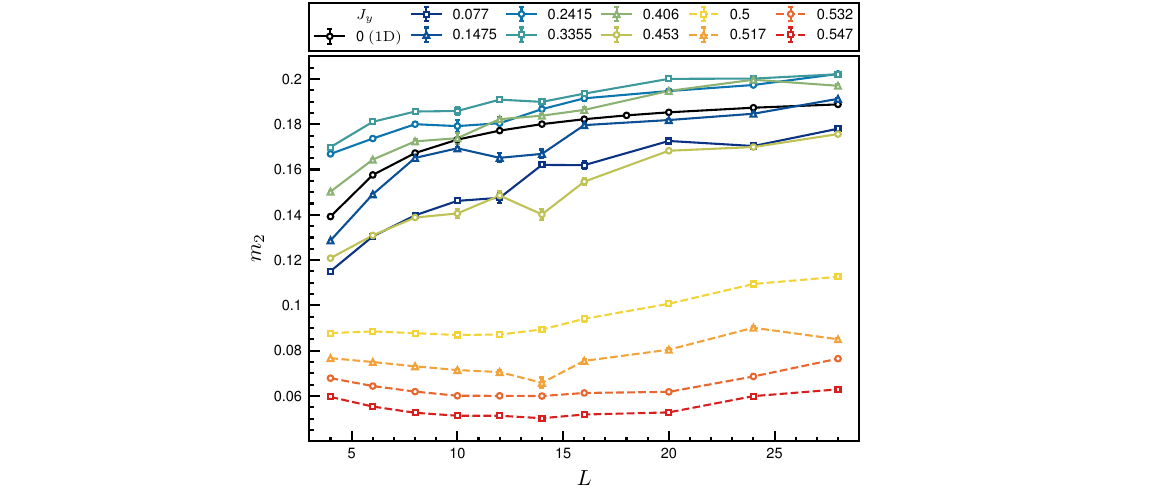}
    \caption{SRE density $m_2$ vs the linear system size $L$. The results were obtained from the same data set used in Fig.~\ref{fig:2}, see the caption for details.}
    \label{fig:10}
\end{figure}
In Fig.~\ref{fig:10} we show the SRE density $m_2 = M_2 / N$ as a function of the linear system size $L$, for different values of the coupling $J_y$. 
Compared with the corresponding $m_1$ data, the estimate of $m_2$ displays visibly larger fluctuations and requires a larger number of samples to reach a comparable stability. Since in our implementation the Majorana string configurations are extracted by perfect sampling from the wave function, this effect is not related to autocorrelation or Markov-chain errors. The relevant source of uncertainty is instead the finite number of independent samples used to estimate the quantity entering the logarithm. 

The difference between the two R\'eny indices $n=1$ and $n=2$ lies essentially in taking an average of a logarithm (in the former) and the logarithm of an average (in the latter). For $n=1$, the estimator can be written schematically as a direct sample average of $\log(\pi_{\rho}(\mb{c}))$; therefore, the finite-sample estimator is unbiased for the corresponding expectation value, up to the usual statistical fluctuations. By contrast, for $n=2$, one first estimates the quantity inside the logarithm. Denoting it by
$Z_2 =\mathbb E[\pi_\rho(\mathbf c)]$, the natural estimator is
\begin{equation}
\widehat Z_2=\frac{1}{N_s}\sum_{s=1}^{N_s}\pi_\rho(\mathbf c_s),
\qquad \mathbf c_s\sim \pi_\rho .
\end{equation}
Although $\widehat Z_2$ is an unbiased estimator of $Z_2$, the estimator of the entropy density, 
\begin{equation}
\widehat m_2 = -\frac{1}{N_{\rm p}} \log \widehat Z_2 - \log 2,
\end{equation}
is not unbiased, because the logarithm is applied after the finite-sample
average has been performed. Since $-\log(x)$ is convex, $\mathbb E[\widehat m_2]\geq m_2$. Thus, the finite-sample estimate of $m_2$ has a systematic upward bias, even when the estimate of $Z_2$ itself is unbiased.  

Expanding around the exact value $Z_2$, one obtains the leading finite-sample bias
\begin{equation}
\mathbb E[\widehat m_2]-m_2
\simeq
\frac{1}{2N_{\rm p}N_s}
\frac{{\rm Var}[\pi_\rho(\mathbf c)]}{Z_2^2}
+O(N_s^{-2}), 
\end{equation}
while the leading variance is
\begin{equation}
{\rm Var}[\widehat m_2] \simeq
\frac{1}{N_{\rm p}^2N_s}
\frac{{\rm Var}[\pi_\rho(\mathbf c)]}{Z_2^2},
\end{equation}
thus showing that the relevant control parameter is not simply the
number of samples, but the relative variance ${\rm Var}[\pi_\rho(\mathbf c)]/Z_2^2$. If this relative variance increases with the system size, the convergence of $m_2$ becomes slow even when the samples are ``perfect'' and ``independent''. This explains why the data in Fig.~\ref{fig:10} are more unstable than the corresponding $m_1$ estimates.

For the same reason, the error bars associated with $m_2$ should be
interpreted with some care. A naive propagation of the uncertainty of $\widehat Z_2$ through the logarithm gives
\begin{equation}
\delta m_2 \simeq
\frac{1}{N_{\rm p}}
\frac{\delta Z_2}{Z_2},
\end{equation}
but this approximation assumes that the fluctuations of $\widehat Z_2$ are small and approximately Gaussian. However, when the distribution is broad or skewed, the logarithmic transformation makes the resulting error bars less reliable and does not capture the finite-sample bias described above.
\end{appendix}

%%%%%%%%% END TODO: CONTENTS

%%%%%%%%%% TODO: BIBLIOGRAPHY
% Provide your bibliography here. You have two options:

%%% FIRST OPTION
% Write your entries here directly, following the example below, including:
% Author(s), Title, Journal Ref. with year in parentheses at the end, followed by the DOI number.

%\begin{thebibliography}{99}
%\bibitem{1931_Bethe_ZP_71} H. A. Bethe, {\it Zur Theorie der Metalle. i. Eigenwerte und Eigenfunktionen der linearen Atomkette}, Zeit. f{\"u}r Phys. {\bf 71}, 205 (1931), \doi{10.1007\%2FBF01341708}.
%\bibitem{arXiv:1108.2700} P. Ginsparg, {\it It was twenty years ago today... }, \url{http://arxiv.org/abs/1108.2700}.
%\end{thebibliography}

%%% SECOND OPTION
% Use your bibtex library, formatted by the SciPost style file.
\bibliography{SciPost_BiBTeX_File.bib}

%%%%%%%%%% END TODO: BIBLIOGRAPHY

\end{document}